\documentclass[11pt]{article}

\usepackage{mathtools, latexsym, color, graphicx, amsmath, amssymb, amsthm}
\usepackage{microtype}
\usepackage{comment}

\def\reals{{\mathbb R}}
\def\eps{{\varepsilon}}
\def\bd{{\partial}}

\def\DDelta{{\boldsymbol{\Delta}}}

\def\A{{\cal A}}

\def\F{{\cal F}}

\def\K{{\cal K}}
\def\L{{\cal L}}
\def\N{{\cal N}}

\def\T{{\cal T}}
\def\H{{\cal H}}
\def\N{{\cal N}}
\def\W{{\cal W}}

\DeclareMathOperator{\polylog}{polylog}

\newtheorem{theorem}{Theorem}[section]
\newtheorem{lemma}[theorem]{Lemma}

\newtheorem{corollary}[theorem]{Corollary}

\title{Intersection Searching amid Tetrahedra in Four Dimensions\thanks{%
  Work by Esther Ezra was partially supported by NSF CAREER under Grant CCF:AF-1553354 
  and by Grant 824/17 from the Israel Science Foundation.
  Work by Micha Sharir was partially supported by Grant 260/18 from the Israel Science Foundation.}}

\author{
Esther Ezra\thanks{%
     School of Computer Science, Bar Ilan University, Ramat Gan, Israel;
     {\sf ezraest@cs.biu.ac.il},
     {https://orcid.org/0000-0001-8133-1335}}
\and
Micha Sharir\thanks{%
     School of Computer Science, Tel Aviv University, Tel Aviv, Israel;
     {\sf michas@tauex.tau.ac.il},
     {http://orcid.org/0000-0002-2541-3763}}
}

\begin{document}

\let\Horig\H

% \begin{titlepage}

\maketitle
\thispagestyle{empty}

\begin{abstract}
We develop data structures for intersection queries in four dimensions that
involve segments, triangles and tetrahedra. Specifically, we study three main problems: 
(i) Preprocess a set of $n$ tetrahedra in $\reals^4$ 
into a data structure for answering segment-intersection queries amid the given tetrahedra 
(referred to as \emph{segment-tetrahedron intersection queries}).
(ii) Preprocess a set of $n$ triangles in $\reals^4$ into a data structure that supports
triangle-intersection queries amid the input triangles (referred to as 
\emph{triangle-triangle intersection queries}).
(iii) Preprocess a set of $n$ segments in $\reals^4$ into a data structure that supports
tetrahedron-intersection queries amid the input segments (referred to as 
\emph{tetrahedron-segment intersection queries}).
In each problem we want either to detect an intersection, or to count or report all intersections.
As far as we can tell, these problems have not been previously studied. 

For problem (i), we first present a ``standard'' solution which, for any prespecified value
$n \le s \le n^6$ of a so-called storage parameter $s$, yields a data structure
with $O^*(s)$ storage and expected preprocessing, which answers an intersection 
query in $O^*(n/s^{1/6})$ time (here and in what follows, the $O^*(\cdot)$ notation 
hides subpolynomial factors). For problems (ii) and (iii), using similar arguments, 
we present a solution that has the same asymptotic performance bounds.

We then improve the solution for problem (i), and present a more intricate data 
structure that uses $O^*(n^{2})$ storage and expected preprocessing, and answers a 
segment-tetrahedron intersection query in $O^*(n^{1/2})$ time. Using the parametric 
search technique of Agarwal and Matou\v{s}ek~\cite{AM:ray}, we can obtain data structures
with similar performance bounds for the \emph{ray-shooting} problem amid tetrahedra in $\reals^4$.
Unfortunately, so far we do not know how to obtain a similar improvement for problems (ii) and (iii).
% \micha{Check!}

Our algorithms are based on a primal-dual technique for range searching with semi-algebraic sets,
based on recent advances in this area~\cite{AAEZ,MP}. As this is a result of independent interest,
we spell out the details of this technique.

We present several applications of our techniques, including continuous collision detection 
amid moving tetrahedra in 3-space, an output-sensitive algorithm for constructing the 
arrangement of $n$ tetrahedra in $\reals^4$, and an output-sensitive algorithm for constructing 
the intersection or union of two or several nonconvex polyhedra in $\reals^4$.
\end{abstract}

% \end{titlepage}

%-------------------------------------------
\section{Introduction}

In this paper we consider various intersection problems involving segments, 
triangles and tetrahedra in $\reals^4$. In four dimensions, the interesting
setups involve (i) intersections between (one-dimensional) query segments and 
(three-dimensional) input tetrahedra, (ii) intersections between (two-dimensional) 
query triangles and (two-dimensional) input triangles, and (iii) intersections 
between (three-dimensional) query tetrahedra and (one-dimensional) input segments.
We study all three problems, and derive efficient solutions to each of them. 

As an interesting application, we consider the \emph{continuous collision detection} problem,
where the input consists of $n$ tetrahedra in $\reals^3$, each of which is moving at some 
constant velocity of its own, and the goal is to detect whether any pair of them collide.
Adding the time as a fourth coordinate, this becomes a batched version of intersection
detection in $\reals^4$, involving both setups (i) (or (iii)) and (ii).
% We may also assume that some of the prisms are bounded, allowing tetrahedra to pop up and disappear 
% (as in a simulation or a video game).
% The goal is to detect a collision between any pair of moving tetrahedra.
% A collision can occur when a vertex $v$ of one tetrahedron hits a face $f$
% of another, or when an edge $e$ of one tetrahedron hits an edge $e'$ of
% another. In the four-dimensional space-time, the first event corresponds 
% to an intersection between the ray traced by $v$ and the three-dimensional 
% prism traced by $f$. The second event corresponds to an intersection between 
% the two-dimensional strip traced by $e$ and the strip traced by $e'$. Again,
% we allow these rays, strips and prisms to be bounded.
% The former (resp., latter) kind of intersections are instances of setup (i) (resp., (ii)).
% 
% In addition to the continuous collision detection application,
% the segment-tetrahedron and triangle-triangle intersection detection problems arise
Other applications include output-sensitive construction of the arrangement of $n$ tetrahedra in $\reals^4$,
and an output-sensitive algorithm for computing the intersection or the union of two or several 
not necessarily convex polyhedra in $\reals^4$. In the three-dimensional versions of these problems, 
which were recently studied by the authors in~\cite{trishoot}, the only setup that needed to be considered
was segment intersection amid triangles. In four dimensions, though, we also face
the triangle-triangle intersection problem, since we also need to find intersections between 
pairs of 2-faces of the input objects. 

%---------------------------------------------------------
\paragraph{Setup (i): Segment-tetrahedron intersection queries.}

Consider first the case of query segments vs.~input tetrahedra.
In the setup considered here, the input objects are $n$ (not necessarily 
disjoint) tetrahedra in $\reals^4$ and the query objects are segments, and 
the goal is to detect, count, or report intersections between the query segment
and the input tetrahedra.

As far as we can tell, this problem has not been explicitly studied so far.
We first present, in Section~\ref{app:trad}, a ``traditional'' (albeit novel) solution, 
in which the problem is reduced to a range searching problem in a suitable parametric 
space, which, in the case of (lines supporting) segments in $\reals^4$, is six-dimensional.
We carefully adapt and combine recent techniques, developed by Agarwal et al.~\cite{AAEZ} 
and Matou\v{s}ek and Pat\'akov\'a~\cite{MP}, which provide algorithmic constructions
of intricate space decompositions based on partitioning polynomials. Using this
machinery, we solve the problem so that, with a so-called storage parameter $s$,
a segment intersection query can be answered in\footnote{%
  As in the abstract, the $O^*(\cdot)$ notation hides subpolynomial factors, typically of the
  form $n^\eps$, for any $\eps > 0$, and their coefficients which depend on $\eps$.}
$O^*(n/s^{1/6})$ time, for any $n \le s \le n^6$, and the storage and preprocessing cost are both $O^*(s)$.

A special case of this setup is an extension to four dimensions of the classical \emph{ray shooting} 
problem, which has mostly been studied in two and three dimensions. In a 
general setting, we are given a collection $S$ of $n$ simply-shaped objects, 
and the goal is to preprocess $S$ into a data structure that supports efficient 
ray shooting queries, where each query specifies a ray $\rho$ and asks for the 
first object of $S$ hit by $\rho$, if such an object exists.
In this work we only consider the (already challenging) case of input tetrahedra.
Using the parametric search technique of Agarwal and Matou\v{s}ek~\cite{AM:ray},
ray shooting queries can be reduced to segment-intersection detection queries,
up to a polylogarithmic factor in the query cost. By the above discussion, we 
obtain the following result:
%---------------------------------
\begin{theorem}
  \label{thm:standard}
  Given a collection $\T$ of $n$ tetrahedra in $\reals^4$, and any storage parameter
  $n\le s\le n^6$, we can preprocess $\T$ into a data structure of size $O^*(s)$, in 
  randomized $O^*(s)$ expected time, so that we can answer any segment-intersection
  or ray-shooting query in $\T$ in $O^*(n/s^{1/6})$ time. The query time bound applies
  to segment-intersection detection and counting queries (and to ray shooting queries).
  The cost is $O^*(n/s^{1/6}) + O(k)$ for reporting queries.
\end{theorem}
%---------------------------------

We later improve upon this standard algorithm in Section~\ref{sec:shoot}, where we show:
%--------------------------------------
\begin{theorem}
  \label{thm:main}
  A collection $S$ of $n$ arbitrary tetrahedra in ${\reals}^4$ can be preprocessed
  into a data structure of size $O^*(n^2)$, in expected time $O^*(n^2)$, which supports
  segment-intersection detection and counting queries and ray-shooting queries in time 
  $O^*(n^{1/2})$ per query.
\end{theorem}
%--------------------------------------
This indeed improves the bounds stated in Theorem~\ref{thm:standard}, which, with $s=O^*(n^2)$ 
storage, has query time $O^*(n^{2/3})$. Furthermore, with the storage bound specified in 
Theorem~\ref{thm:main}, the query bound is similar to that obtained for ray-shooting amid 
hyperplanes (rather than tetrahedra) in ${\reals}^4$~\cite{AM:ray}.

\begin{figure}[htb]
  \begin{center}
    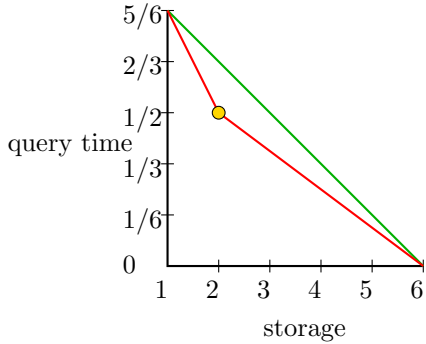
    \caption{{\sf The tradeoff between storage and query time. The breakpoint in the graph 
    represents the case studied in Theorem~\ref{thm:main}. Both axes are drawn on a logarithmic scale.}}
    \label{fig:trade}
  \end{center}
\end{figure}

We then go on to extend the result to obtain a tradeoff between storage (and expected 
preprocessing time) and query time. We obtain the following result.
%---------------------------------
\begin{theorem} \label{thm:tradeoff}
Let $\T$ be a set of $n$ tetrahedra in $\reals^4$.
With storage parameter $s$, which can vary between $n$ and $n^6$, we can answer 
a segment intersection or a ray shooting query amid the tetrahedra of $\T$ in time
%-----------------------
\begin{equation}
  \label{eq:trade1}
  Q(n,s) =
  \begin{cases}
    O^*\left(\frac{n^{7/6}}{s^{1/3}} \right) & \text{for } s = O(n^2) \\
    O^*\left(\frac{n^{3/4}}{s^{1/8}} \right) & \text{for } s = \Omega(n^2) .
  \end{cases}
\end{equation}
%-----------------------
Again, this bound pertains to detection and counting queries, and incurs the additional
term $O(k)$ for reporting queries, where $k$ is the output size.
\end{theorem}
%--------------------------------
See Figure~\ref{fig:trade} for an illustration. This implies the following corollary.
%---------------------------------
\begin{corollary}
  \label{cor:queries}
  One can answer $m$ segment intersection detection or counting queries, 
  or ray-shooting queries, on $n$ tetrahedra in $\reals^4$ in 
  %-----------------------
  \begin{equation}
    \label{eq:trade2}
  \max \left\{ O^*(m^{3/4}n^{7/8} + n),\; O^*(m^{8/9}n^{2/3} + m) \right\}
  \end{equation}
  %-----------------------
  time and storage. The first (resp., second) bound dominates when $m \le n^{3/2}$ 
  (resp., $m \ge n^{3/2}$).
\end{corollary} 
%---------------------------------

%---------------------------------------------------------
\paragraph{Setup (ii): Triangle-triangle intersection queries.}

We next consider the second setup of intersection queries, where both 
input and query objects are triangles in $\reals^4$.
We show that this setup can also be reduced, similar to setup (i),
to a multi-level range searching problem in $\reals^6$ involving 
semi-algebraic ranges. This allows us to obtain the same performance 
bounds here too. Namely we have:
%---------------------------------
\begin{theorem}
  \label{thm:qsetup2}
  Given a collection $\DDelta$ of $n$ triangles in $\reals^4$, and any storage parameter
  $n\le s\le n^6$, we can preprocess $\DDelta$ into a data structure of size $O^*(s)$, in 
  randomized $O^*(s)$ expected time, so that we can answer any triangle-intersection
  query in $\DDelta$ in $O^*(n/s^{1/6})$ time.
\end{theorem}
%---------------------------------

Since both input and query objects are triangles, it is also interesting to
consider the bichromatic batched version of the problem. Namely we have:
%---------------------------------
\begin{theorem}
  \label{thm:setup2}
  Given two collections $R$ and $B$ of triangles in $\reals^4$, of respective sizes $m$ and $n$,
  We can detect an intersection between some triangle of $R$ and some triangle of $B$, or count
  all such intersections, in time $O^*(m^{6/7}n^{6/7} + m + n)$. We can also report all these
  intersections in time $O^*(m^{6/7}n^{6/7} + m + n + k)$, where $k$ is the output size.
\end{theorem}
%---------------------------------

As a consequence, integrating this bound with the one obtained in Theorem~\ref{thm:standard} 
(in which we need to set $s = n^{12/7}$ to match the performance with that stated above, 
as is easily verified), we obtain an overall $O^*\left(n^{12/7}\right)$ expected-time 
solution for the continuous collision detection problem, that is:\footnote{%
  Here we use an obvious divide-and-conquer approach in order to reduce the general 
  (non-bichromatic) problem to the bichromatic version.}
%---------------------------------
\begin{theorem}
  \label{thm:collision}
  Given $n$ tetrahedra in $\reals^3$, each of which is moving at some constant velocity 
  of its own, one can detect a collision between any pair of moving tetrahedra in 
  $O^*\left(n^{12/7}\right)$ expected time. 
\end{theorem}
%---------------------------------

Collision detection has been widely studied---see Lin, Manocha and Kim~\cite{LMK} for
a recent comprehensive survey, and the references therein. We are not aware of any work 
that addresses the exact algorithmic approach for the specific setup considered here, 
although there are some works, such as Canny~\cite{Can} or Sch\"omer and Thiel~\cite{ST-95},
that address similar contexts.

We then consider the applications of our techniques to the problems of output-sensitive construction 
of an arrangement of tetrahedra in $\reals^4$, and of constructing the intersection or union
of two or several (nonconvex) polyhedra in $\reals^4$. Using the bounds for setups (i) and (ii), we obtain,
in Section~\ref{sec:arr}:
% \micha{General position has just popped up. Resolve.}
%-----------------------------------
\begin{theorem}
  \label{thm:arr}
  (i) Let $\T$ be a collection of $n$ tetrahedra in general position in $\reals^4$.
  We can construct the arrangement $\A(\T)$ of $\T$ in
  $O^*(n^{12/7} + n^{1/2}k_2 + k_4)$ randomized expected time, where $k_2$ is the number of 
  intersecting pairs of tetrahedra in $\T$, and $k_4$ is the number of 
  vertices of $\A(\T)$. 
  (ii) Given two arbitrary polyhedra $R$ and $B$ in $\reals^4$, each of complexity $O(n)$
  (where the complexity is the number of faces of all dimensions on their boundary),
  that lie in general position with respect to one another,
  the intersection $R \cap B$ can be computed in expected time $O^*(n^{12/7} + n^{1/2}k_2 + k_4)$,
  where $k_2$ is the number of 2-faces of $\A(R \cup B)$, and $k_4$ is the number of 
  vertices of $\A(R \cup B)$. 
  %\esther{Shall we make a general position assumption here? say that if a tetrahedron of $R$ intersects a tetrahedron of $B$ then they intersect in a two-dimensional convex polygon.}
\end{theorem}
%-----------------------------------

As another application of our technique we present an efficient algorithm for detecting,
counting or reporting intersections between $n$ 2-flats and $n$ lines in $\reals^4$. 
We show that, given $n$ lines and $n$ 2-flats in $\reals^4$, one can detect whether
some line intersects some 2-flat in $O^*(n^{13/8})$ expected time, or count the number
of such intersections. One can also report all $k$ intersections in $O^*(n^{13/8} + k)$ 
expected time. This result is a degenerate special case of the triangle-triangle intersection 
setup (ii), and admits a faster solution. (Note that in general position 2-flats and lines 
are not expected to meet in $\reals^4$, which makes this special case interesting.)

%------------------------
\paragraph{Setup (iii): Tetrahedron-segment intersection queries.}
This is a symmetric version of setup (i), where the input consists of $n$ segments in 
$\reals^4$ and the query is with a tetrahedron $T$, where the goal is to detect, count 
or report intersections between $T$ and the input segments. Using a similar machinery, 
we obtain the same asymptotic performance bounds, as in the standard solutions, for this setup too.
%--------------
\begin{theorem}
  \label{thm:standard_dual}
  Given a collection $S$ of $n$ segments in $\reals^4$, and any storage parameter
  $n\le s\le n^6$, we can preprocess $S$ into a data structure of size $O^*(s)$, in 
  randomized $O^*(s)$ expected time, so that we can answer any tetrahedron-intersection
  query in $S$ in $O^*(n/s^{1/6})$ time. The query time bound applies
  to tetrahedron-intersection detection and counting queries.
  The cost is $O^*(n/s^{1/6}) + O(k)$ for reporting queries.
\end{theorem}
%--------------

The paper is organized as follows. We first present, in Section~\ref{app:trad},
the standard (novel) technique for setup (i). A simple modification of the algorithm,
also presented in Section~\ref{app:trad}, yields an algorithm for Setup (iii).
The algorithm for setup (ii) is
then presented in Section~\ref{sec:bip}. The improved algorithm for setup (i) 
is presented in Section~\ref{sec:shoot}. This improved solution can be extended 
to yield an improved tradeoff between storage and query time, the details of which
are given in Section~\ref{app:trade}. Our applications, for constructing arrangements
of tetrahedra in $\reals^4$, constructing the intersection or union of nonconvex 
polyhedra in $\reals^4$, and continuous collision detection, are presented in
Section~\ref{sec:arr}. Finally, in Section~\ref{app:int}, we study the special case
of intersections between 2-fats and lines in $\reals^4$.

%---------------------------------------
\section{Segment-Intersection amid Tetrahedra: \\ An Initial Algorithm}
\label{app:trad}

% \micha{The presentation in this section is very detailed, including
% details of most of the off-the-shelf tools that we use. I personally don't mind,
% just commenting. A person like Pankaj might be unhappy with such a level of detail.}
% \esther{Now everything is in the appendix.}

In this section we present an initial solution to the problem of segment-intersection 
detection amid tetrahedra in four dimensions, which is based on a careful combination
of the recent range searching machinery of \cite{AAEZ,MP}. 
As far as we can tell, such a solution has not yet been presented in the 
literature. Also, the adaptation of the available techniques to this problem is 
not simple, requires nontrivial and careful enhancements, and is sufficiently novel 
to be of independent interest. Moreover, this gives a yardstick for appreciating the 
improvement obtained by our improved algorithm, presented in Section~\ref{sec:shoot}.
The machinery developed here will be used, with some appropriate modifications,
in the algorithms for handling setups (ii) and (iii).

The parametric search technique of Agarwal and Matou\v{s}ek~\cite{AM:ray}
reduces ray shooting queries to segment-intersection detection queries,
so it suffices to consider the latter problem. The reporting and counting 
variants are simple extensions of the same technique, as will be discussed as we go.
% \esther{At the moment, I do not discuss counting and reporting.}

To obtain a tradeoff between the storage of the structure (and its preprocessing cost)
and the query time,
our algorithm uses a primal-dual approach. However, both the primal and dual setups
suffer from the fact that, in four dimensions, segments and tetrahedra require 
too many parameters to specify. Specifically, a segment requires eight parameters 
(e.g., by specifying its two endpoints), while a tetrahedron requires $16$ 
parameters (e.g., by specifying the coordinates of its four vertices). 

To address this issue, we use a multi-level data structure, where each
level caters to one aspect of the condition that a segment crosses a 
tetrahedron. This is done so that, at each of these levels, the number 
of parameters that a segment or a tetrahedron requires is at most six.

Specifically, the condition that a segment $e$, that lies on a line $\ell$,
intersects a tetrahedron $\Delta$, supported by a hyperplane $h_\Delta$,
is the conjunction of the following conditions:

\medskip
\begin{itemize}
\item[(i)]
The two endpoints of $e$ lie on different sides of $h_\Delta$.

\medskip
\item[(ii)]
With a suitable choice of a direction of $\ell$ and an orientation
of $\Delta$, $\ell$ has a positive orientation with respect to each 
of the 2-planes that support the four 2-faces of $\Delta$.
\end{itemize}

\medskip
Conditions (i) and (ii) are the conjunction of a total of six sub-conditions: 
each of the first two conditions tests the position of some endpoint of $e$ 
with respect to the hyperplanes $h_\Delta$, and each of the other four conditions
tests the orientation of $\ell$ with respect to the 2-planes supporting specific 
2-faces of the tetrahedra. Thus, the dual structure 
has six levels, two for testing the sub-conditions of condition (i) 
and four for testing the sub-conditions of condition (ii). 

More precisely, each but the last level is a collection of structures, 
each of which operates on some canonical subset of tetrahedra, produced at
the previous levels, and collects all the tetrahedra $\Delta$ of that subset 
that satisfy the corresponding sub-condition for the query segment (that a 
specific endpoint of $e$ lies in a specific side of $h_\Delta$ for the first 
two levels, or that the oriented 2-plane supporting a specific 2-face 
of $\Delta$ is positively oriented with respect to the directed line $\ell$ 
for the last four levels), as the disjoint union of precomputed canonical sets
of tetrahedra. The last level just tests whether the last sub-condition
is satisfied for any tetrahedron in the current canonical set.

We use the fact that lines in $\reals^4$ require six
real parameters to specify. The space of lines in $\reals^4$ is actually
projective, but for simplicity of presentation we regard it as a real space,
and ignore the special cases in which the real representation fails. 
Handling these cases follows the same approach, and is in fact simpler.
Alternatively, a generic (say random) rotation of the coordinate frame 
allows us to ignore them altogether.

One simple way to represent a line $\ell$ in $\reals^4$ is by the points
$u^0_\ell = (x_0, y_0, z_0, 0)$ and $u^1_\ell = (x_1, y_1, z_1, 1)$ at 
which $\ell$ crosses the hyperplanes $w=0$ and $w=1$, respectively 
(ignoring lines that are orthogonal to the $w$-axis), so the line
$\ell$ can be represented as the point 
$p_\ell = \left( x_0, y_0, z_0, x_1, y_1, z_1 \right)$ in $\reals^6$, as desired.

Similarly, 2-planes in $\reals^4$ also require six parameters to specify.
This is simply because the duality in $\reals^4$ maps lines to 2-planes
and vice versa, but a concrete way to represent 2-planes by six parameters,
which we will use, 
is to specify three points on a 2-plane $\pi$ that are intersections of 
$\pi$ with three fixed 2-planes, such as, say, $x=y=0$, $x=0$ and $y=1$, 
and $x=y=1$ (again ignoring special directions of $\pi$). Each of the 
intersection points has two degrees of freedom (as two of its coordinates 
are fixed), for a total of six. Denote these points as 
$v^{(00)}_\pi$, $v^{(01)}_\pi$, and $v^{(11)}_\pi$, and put
$q_\pi = \left( v^{(00)}_\pi, v^{(01)}_\pi, v^{(11)}_\pi \right)$, 
listing only the $w$- and $z$-coordinates of each point, so $q_\pi$
is indeed a point in $\reals^6$. 

These observations are meaningful only for the last four levels of the 
structure. The first two levels are simpler, as they deal with points
(the endpoints of $e$) and hyperplanes (those supporting the tetrahedra 
of $\T$) in $\reals^4$. Thus each of the first two levels is a halfspace 
range searching structure for points and halfspaces in $\reals^4$.
(Actually, this is the case when we pass to the dual 4-space; in the
primal we have a point-enclosure problem, where the query is a point
and the input consists of halfspaces bounded by the relevant hyperplanes.)
Using standard techniques (see, e.g.,~\cite{Ag:rs}), this can be done,
for $N$ halfspaces in the current canonical subset and using $O^*(N)$ storage, 
so that a query costs $O^*(N^{3/4})$ time.\footnote{%
  A tradeoff between storage and query time is also available, but we
  do not need it here.}
This cost will be subsumed by the query time 
bounds for the last four levels. The cost of a query includes the cost 
of reporting its output, as a list of canonical sets.
% (but not of enumerating the elements of these sets, which is too expensive and not needed anyway).

We next consider the (more involved) situation in the last four levels
of the structure. Here the query segment is replaced by its supporting
line $\ell$, and each tetrahedron $\Delta$ is replaced by the 2-plane 
supporting a specific 2-face of $\Delta$. In the primal setup, the line
$\ell$ is represented as a point $p_\ell$ in (projective) 6-space, in 
the manner just described, and a tetrahedron $\Delta$, represented by 
a suitable 2-plane $\pi$, is represented as a semi-algebraic region 
$K_\pi$, consisting of all points that represent (directed) lines that 
are positively oriented with respect to $\pi$. The problem that we face
is a point-enclosure query, in which we want to collect all the regions
$K_\pi$ that contain $p_\ell$.
In the dual setup, the 2-planes $\pi$ are represented as 
points in $\reals^6$, and the (directed) query line $\ell$ is represented 
as a semi-algebraic region $Q_\ell$ that consists of all (oriented) 
2-planes that are positively oriented with respect to $\ell$. The 
problem here is a semi-algebraic range searching query, where we want 
to collect all the input points in $Q_\ell$.

The orientation test of $\ell$ with respect to $\pi$ amounts to 
computing the sign of the $5\times 5$ determinant
\begin{equation} \label{eq:det}
\begin{vmatrix}
u^0_\ell & 1 \\
u^1_\ell & 1 \\
v^{(00)}_\pi & 1 \\
v^{(01)}_\pi & 1 \\
v^{(11)}_\pi & 1 
\end{vmatrix} ,
\end{equation}
with a suitable orientation of the pair of points $u^0_\ell$, $u^1_\ell$ 
on $\ell$ (dictating the direction of $\ell$), and of the triple of 
points $v^{(00)}_\pi$, $v^{(01)}_\pi$, $v^{(11)}_\pi$ on $\pi$
(dictating the orientation of $\pi$).

To compute these signs, at each of the four latter levels of the structure, 
we use a primal-dual approach, where the top part of the structure is in 
the primal, and at each of its leaf nodes we pass to the dual.

%----------------------------------
\paragraph{The dual setup.}
The dual setup is simpler, so we begin with its description.
In the dual setup, each tetrahedron $\Delta$ of the current canonical 
subset of $\T$ is mapped to the point
$q_\pi = \left( v^{(00)}_\pi,\, v^{(01)}_\pi,\, v^{(11)}_\pi \right)$
in $\reals^6$, where $\pi$ is the 2-plane supporting the 2-face of $\Delta$ 
that corresponds to the present level. As just mentioned, the query line $\ell$ is 
mapped to a semi-algebraic region $Q_\ell$ of constant complexity in $\reals^6$, 
consisting of all points $q_\pi$ that represent (oriented) 2-planes that 
have positive orientation with respect to $\ell$, that is, the corresponding
determinant in (\ref{eq:det}) is positive. ($Q_\ell$ is in fact defined by
a single polynomial inequality, where the polynomial is cubic in $q_\pi$.)

As already mentioned, the task at hand, at each but the last level, is 
to collect the points $q_\pi$ that lie in $Q_\ell$, as the disjoint 
union of a small number of precomputed canonical sets of tetrahedra, and
the task at the last level, for detection queries, is to determine whether 
$Q_\ell$ contains any point $q_\pi$, for $\pi$ corresponding to the last 
2-faces of the tetrahedra in the present canonical subset of $\T$. For
counting queries, we add the size of each output canonical set to a global 
counter, and for reporting queries we output the elements of each output set.
In other words, we have, at each of these levels, a problem involving range 
searching with semi-algebraic ranges in $\reals^6$. Using the algorithm of 
Matou\v{s}ek and Pat\'akov\'a~\cite{MP}, which is a simplified version 
of the algorithm of Agarwal et al.~\cite{AMS}, this can be done, for $N$ 
tetrahedra with $O^*(N)$ storage, so that a detection or counting query 
takes $O^*(N^{5/6})$ time (including the cost of reporting, without 
enumerating, the output canonical sets). Reporting queries are handled
and analyzed in a suitably modified manner. 
See \cite[Theorem~6.1]{Ag:rs} for more details.

%----------------------------------
\paragraph{The primal setup.}
With this procedure at hand, we go back to the primal structure, at 
each of the last four levels. As noted, the problem that we face there 
is a point enclosure problem, where the input consists of some $N$
constant-complexity semi-algebraic regions in $\reals^6$ of the form 
$K_\pi$, and the query is the point $p_\ell$ that represents $\ell$, 
as defined earlier, and the task is to collect all the regions $K_\pi$ 
that contain $p_\ell$, as the disjoint union of a small number of 
precomputed canonical sets, or, at the last level,
to determine whether $p_\ell$ is contained in any such region.
Here $K_\pi$ is given by a single polynomial inequality, and the
polynomial is quadratic in $p_\ell$.

This problem has recently been studied in Agarwal et al.~\cite{AAEZ},
using a multi-level polynomial partitioning technique, for 
the case where we allow maximum storage for the structure (that is, 
$O^*(N^6)$ in our case) and want the query time to be logarithmic.
We next show that the structure can be modified so that its preprocessing
stops `prematurely' when its overall storage attains some prescribed value,
and each of the subproblems at the new leaves can be handled via the 
dual algorithm presented above.

The crucial technical tool in \cite{AAEZ}, on which their technique 
is based, is the following result. We give here a restricted specialized
version that suffices for our purposes.
(When applying this tool in $d$ dimensions, the parameter $6$ has to be
replaced by $d$.)

%--------------------------------
\begin{theorem}[A specialized version of Agarwal et al.~\protect{\cite[Corollary 4.8]{AAEZ}}] \label{thm:aaez}
Given a set $\Psi$ of $N$ constant-degree algebraic surfaces in $\reals^6$, 
and a parameter $0 < \delta < 1/6$,
there are finite collections $\Omega_0,\ldots,\Omega_6$ of semi-algebraic 
sets in $\reals^6$ with the following properties.
\begin{itemize}
\item
For each index $i$, each cell $\omega\in \Omega_i$ is a connected semi-algebraic 
set of constant complexity. The size $|\Omega_i|$ of $\Omega_i$ (the number of its sets) 
is a constant that depends on $\delta$.
\item
For each index $i$ and each $\omega\in \Omega_i$, at most 
$\frac{N}{4|\Omega_i|^{1/6-\delta}}$ surfaces from $\Psi$ cross $\omega$ 
(intersect $\omega$ without fully containing it).
\item
The cells partition $\reals^6$, in the sense that
${\displaystyle \reals^6 = \bigsqcup_{i=0}^6 \bigsqcup_{\omega\in \Omega_i} \omega}$,
where $\bigsqcup$ denotes disjoint union.
\end{itemize}
The sets in $\Omega_0,\ldots,\Omega_6$ can be computed in $O(n + m)$ expected time,
where the constant of proportionality depends on $\delta$, by a
randomized algorithm. For each $i$ and for every set $\omega\in \Omega_i$, the 
algorithm returns a semi-algebraic representation of $\omega$, a reference 
point inside $\omega$, and the subset of surfaces of $\Psi$ that cross $\omega$.
\end{theorem}
%--------------------------------

In our case, the surfaces of $\Psi$ are the boundaries of the regions $K_\pi$
(each defined by a single quadratic polynomial equation).
A straightforward enhancement of the algorithm of \cite{AAEZ} also yields, for 
each $i$ and each $\omega\in\Omega_i$, the set of regions $K_\pi$ that fully 
contain $\omega$, within the same asymptotic time bound. 

We compute the partition of Theorem~\ref{thm:aaez} and find, for each 
$\psi = \bd K_\pi\in\Psi$, the sets $\omega \in \Omega_i$, over all 
$i=0,\ldots,6$, that $\psi$ crosses, and those that are fully contained in
$K_\pi$. For each $i$ and $\omega \in \Omega_i$, let $\K_{i,\omega}$ 
(resp., $\K^0_{i,\omega}$) denote the set of tetrahedra $\Delta\in \T$ 
for which $\bd K_\pi$ crosses $\omega$ (resp., $K_\pi$ fully contains $\omega$).

The overall size of the sets $\K^0_{i,\omega}$, over all $i$ and 
$\omega\in \Omega_i$, is $O(N)$, with a constant that depends on $\delta$
(that is, on the sizes $|\Omega_i|$, which depend on $\delta$).
% \micha{We should make it explicit that the $|\Omega_i|$'s are indeed of constant size.}

For each $i$ and $\omega$ we also have a recursive subproblem that involves
the subset $\K_{i,\omega}$ of the tetrahedra $\Delta$ for which $\bd K_\pi$ 
crosses $\omega$. Putting $r_i := |\Omega_i|$, for $i=0,\ldots,6$, we have, 
for each $i$ and $\omega$,
${\displaystyle |\K_{i,\omega}| \le \frac{N}{4r_i^{1/6-\delta}}}$.
We run the recursion, but not all the way through, as in \cite{AAEZ}.
Instead, we use the following storage allocation rule. We fix the storage
that we are willing to allocate to the structure, and distribute it
among the nodes of the recursion, as follows. To simplify the analysis,
we distinguish between the storage itself, and the so-called 
\emph{storage parameter} $s$, which is what we actually manage, but
we have the property that the actual storage will always be $O^*(s)$.

Let $s$ be the storage parameter that we allocate at the root of the
structure. For each $i$ and each set $\omega\in\Omega_i$, we allocate
the storage parameter $s/(4|\Omega_i|)$ for $\omega$. Hence, when we 
reach some set $\omega$ at a deeper level of recursion, say level $j$,
the storage parameter allocated to $\omega$ is
${\displaystyle \frac{s}{4^j |\Omega_{i_1}^{(1)}|\cdot |\Omega_{i_2}^{(2)}| \cdots |\Omega_{i_j}^{(j)}| }}$,
where $\Omega_{i_1}^{(1)},\, \Omega_{i_2}^{(2)},\ldots, \Omega_{i_j}^{(j)}$,
for indices $0\le i_1, i_2, \ldots, i_j \le 6$, are the partition families 
at the ancestors of $\omega$ in the recursion.

We stop the recursion when we reach nodes for which the allocated storage 
parameter is (roughly) equal to the number of tetrahedra at the node;
a more precise statement of the termination rule is given shortly.

Put, for each set $\omega$,
${\displaystyle r_\omega := |\Omega_{i_1}^{(1)}|\cdot |\Omega_{i_2}^{(2)}| \cdots |\Omega_{i_j}^{(j)}|}$,
using the above notation for $\omega$. The storage parameter allocated to
$\omega$ is thus $s/(4^jr_\omega)$. Also, by Theorem~\ref{thm:aaez}, the number of
tetrahedra $\Delta$ that participate in the subproblem at $\omega$ is at most
\[
\frac{n}{4^j|\Omega_{i_1}^{(1)}|^{1/6-\delta}\cdot |\Omega_{i_2}^{(2)}|^{1/6-\delta} \cdots |\Omega_{i_j}^{(j)}|^{1/6-\delta} } 
= \frac{n}{4^jr_\omega^{1/6-\delta}} ,
\]
and the stopping condition that we use is that
${\displaystyle \frac{s}{4^jr_\omega} = \frac{n}{4^jr_\omega^{1/6-\delta}}}$, or 
${\displaystyle r_\omega = (s/n)^{(6/5)/(1+6\delta/5)}}$.
The size of a subproblem at a leaf is 
(using the $O^*(\cdot)$ notation to hide exponents that are proportional to $\delta$
and constants of proportionality that depend on $\delta$)
\[
n_\omega = \frac{n}{4^jr_\omega^{1/6-\delta}} = \frac{1}{4^j} O^*\left( \frac{n^{6/5}}{s^{1/5}} \right) %. 
= O^*\left( \frac{n^{6/5}}{s^{1/5}} \right) . 
\] 
In more detail, since all the parameters $r_j$
in Theorem~\ref{thm:aaez} are at least some sufficiently large constant that
we can control, we can make the factor $4^j$ to be $O(s^\delta)$, for any $\delta>0$ of our choice.
To be more precise, the choice of $\delta$ determines how large the parameters $r_j$
have to be taken to ensure that $4^j = O(s^\delta)$, and the choice of these parameters
adds a constant factor to the query cost (incurred by the cost of locating, in brute 
force, the cells $\omega$, at the various recursive levels, that contain $p_\ell$),
which depends on these parameters, and thus on $\delta$.

At each leaf $\omega$ we pass to the dual structure reviewed above. 
It uses $O^*(n_\omega)$ storage and answers a query in time 
$O^*(n_\omega^{5/6}) = O^*(n/s^{1/6})$. 
To answer a query with a line $\ell$ in the combined structure, 
we search with its point $p_\ell$ in the primal substructure, in 
$O(\log n)$ time (with a constant of proportionality that depends on $\delta$; 
see below), to locate the leaf cell $\omega$ that contains $p_\ell$ 
(using the properties that (a) each recursive step involves a partitioning of 
constant size, and (b) the cells in the partition are pairwise disjoint).
We then search with $Q_\ell$ in the dual structure at $\omega$, which takes,
as just noted, $O^*(n/s^{1/6})$ time. The overall cost of the 
query is therefore $O^*(n/s^{1/6})$.

As to the actual storage used by the structure, the allocation mechanism ensures
that each level of the recursion uses storage that is at most $7/4$ times larger
than the storage used in the previous level, because each node has seven 
child collections $\Omega_0, \ldots, \Omega_6$, each of which is allocated 
an amount of storage $s/4$. Hence the overall storage used is $O((7/4)^j s)$, 
where $j$ is the recursion depth. Arguing as in the query time analysis,
we can make the factor $(7/4)^j$ to be $O(s^\delta)$, for any $\delta>0$.
That is, the overall storage used is $O(s^{1+\delta})$, or, in our notation, $O^*(s)$. 

The above description of the structure applies to any single level among the
four latter levels of the structure. The first two levels are considerably 
simpler and more efficient. The primal-dual approach is straightforward for 
halfspace range searching, and the parametric dimension is only four for the
first two levels. The standard machinery (reviewed, e.g., in \cite{Ag:rs}) 
implies that, with $s$ storage and $N$ input tetrahedra, the cost of a 
query at each of these levels is $O^*(N/s^{1/4})$.

Putting everything together, and using standard arguments in the analysis 
of multi-level structures (see \cite[Theorem~6.1]{Ag:rs} for details), the 
overall size of the six-level structure is $O^*(s)$, for any prescribed 
storage parameter $s$ between $n$ and $n^6$, and a query takes $O^*(n/s^{1/6})$ 
time. That is, this finally concludes the proof of Theorem~\ref{thm:standard}
for the case of intersection detection queries. Counting and reporting queries
are handled similarly, with a similar analysis, exploiting the fact that the
decomposition in Theorem~\ref{thm:aaez} is into disjoint subsets, as is a similar 
decomposition used in the machinery of \cite{MP}. For reporting queries, their cost
involves an additional term $O(k)$, where $k$ is the output size.
$\Box$

% \micha{Look also at my comment below. Maybe in view of that we should just remove this part?}

\medskip
\noindent{\bf Remark.}
Our mechanism is in fact a special instantiation of the following general result, 
which is of independent interest, and which yields a trade-off bound for 
semi-algebraic range searching in any dimension $d$. That is, consider a general 
problem of this kind, that involves $n$ points in $\reals^d$, and aims to answer 
semi-algebraic range queries, where the ranges have constant complexity, and each
range has $d$ degrees of freedom (so the problem has a symmetric dual version).
One can solve such a problem in time $O^*(n/s^{1/d})$ per query, using $O^*(s)$ 
space and preprocessing, where $s$ is any parameter between $n$ and $n^d$. These 
queries include detecting whether a query range contains any input point,
counting the number of such points, or reporting them (with an additional term $O(k)$ 
in the query cost, where $k$ is the output size).
Using duality, we obtain the same performance bounds for point-enclosure queries, 
where the input consists on $n$ constant-complexity semi-algebraic regions in 
$\reals^d$, and the query is with a point $p$, where the goal is to detect, count 
or report containments of $p$ in the input regions.
The same asymptotic bound is obtained for simplex range searching~\cite{Ag:rs}, but
our analysis shows that this bound corresponds to a much more general family of query ranges.
The two extreme cases $s=n$ and $s=n^d$ have been treated in \cite{MP} and \cite{AAEZ},
respectively, but the tradeoff between these extreme cases has not been treated explicitly 
(for $d > 4$), as far as we can tell. As evidenced in the preceding analysis, this tradeoff 
is not as routine as one might think, because of the complicated nature of the partitioning 
used in Theorem~\ref{thm:aaez} (as well as in \cite[Theorem 1.1]{MP}). 
We summarize this result in the following corollary:

% \micha{This generalization is not completely kosher, because it relies on the 
% property that in the primal and in the dual we have the same number of degrees 
% of freedom (for the query and the input objects). In general, when these numbers 
% are not the same, we get `hybrid' exponents that involve both parameters.}

%-----------------------------
\begin{theorem}
  \label{cor:range_search}
  Let $P$ be a set of $n$ points in $\reals^d$, for any dimension $d$,
  and let $\Gamma$ be a family of semi-algebraic ranges of constant complexity 
  in $\reals^d$, each of which has $d$ degrees of freedom. Let $n \le s \le n^d$ 
  be a prespecified storage parameter.
  Then one can preprocess $P$ into a data structure of storage and preprocessing 
  $O^*(s)$, such that a range-query, with a range $\gamma \in \Gamma$, can be 
  answered in $O^*(n/s^{1/d})$ time. Such queries include detecting whether 
  $\gamma$ contains any point of $P$, counting the number of such points, and 
  reporting them (with an additional $O(k)$ term in the latter case, where 
  $k$ is the number of these points). The same performance bounds apply to the 
  dual point-enclosure case, where the input consists of $n$ regions from $\Gamma$ 
  and the query is with a point $p\in\reals^d$.
\end{theorem}
%-----------------------------

% \micha{Remove up to now; the rest is sort of OK.}

\medskip
\noindent{\bf Remarks. (i)}
Theorem~\ref{cor:range_search} can be extended to the case where the number
of degrees of freedom of the ranges is different from $d$, but the resulting
performance bound has a more complicated expression, which is not spelled out
in this work. See, e.g.,~\cite{AAEKS} for a recent work in progress that handles 
the asymmetric setup. 

\medskip
\noindent{\bf (ii)}
We note that our technique can be extended to segment intersection detection
queries amid a collection of $n$ $(d-1)$-simplices in any dimension $d$. In 
that case the structure has $d+2$ levels. The first two levels ensure that the 
endpoints of the query segment $e$ lie on different sides of the hyperplane
containing the input simplex $\Delta$, and are implemented by halfspace range
searching structures in $\reals^d$. The last $d$ levels ensure that the 
line containing $e$ has positive orientation with respect to each of the
$(d-2)$-flats containing the facets of $\Delta$, with suitable orientations
of the line and the flats. Since lines and $(d-2)$-flats in $\reals^d$ have 
$2d-2$ degrees of freedom, these levels are implemented using semi-algebraic 
range searching structures, where both primal and dual parts are in $\reals^{2d-2}$. 
Hence the cost of the query at each of the last $d$ levels dominates the overall 
cost, which is thus $O^*(n/s^{1/(2d-2)})$. The parameter $s$ can vary between 
$n$ and $n^{2d-2}$.

%------------------------------------------
\paragraph{Setup (iii).}
A very similar mechanism, with the same performance bounds, handles the reverse 
situation of setup (iii), in which the input is a set of $n$ segments in $\reals^4$, 
and the query is with a tetrahedron $T$, and the goal is to detect, count, or report
intersections between $T$ and the input segments. The algorithm and its analysis are
very similar to those given above, except that we have to flip the roles of points
and hyperplanes (in the first two levels of the structure) and of lines and 2-planes
(in the last four levels of the structure).

The resulting algorithm is what is asserted in Theorem~\ref{thm:standard_dual}.

%-----------------------------------------------------------
\section{Triangle-Triangle Intersection Queries in $\reals^4$}
\label{sec:bip}

% \paragraph{Preprocessing and query.}
% We first consider the preprocessing-and-query version of the problem.
Let $\DDelta$ be a set of $n$ triangles in $\reals^4$. We consider
various triangle-triangle intersection problems, the simplest 
of which is just to detect whether a query triangle intersects
any triangle of $\DDelta$. Alternatively, we may want to count or to report 
all such intersections. For concreteness we focus on the
detection problem in what follows, but, as in the previous section,
the algorithm can easily be extended to also handle the other kinds of problems.

Similar to the preceding section, we use a multi-level data structure, 
where each level caters to one aspect of the condition that a triangle
crosses another triangle. Specifically, let $\Delta_1$ and $\Delta_2$
be two triangles, and let $\pi_1$, $\pi_2$ be the respective 2-planes
that contain them. Assuming general position, $\pi_1$ and $\pi_2$
always intersect at a single point $\xi$, and $\Delta_1$ intersects
$\Delta_2$ if and only if $\xi$ belongs to both triangles. As is
easily verified, this latter condition is equivalent, with suitable 
orientations of $\pi_1$, $\pi_2$, and of the lines supporting the edges 
of both triangles, to the conjunction of the following conditions:

\medskip
\begin{itemize}
\item[(i)]
$\pi_1$ is positively oriented with respect to each of the lines
that support the edges of $\Delta_2$.

\medskip
\item[(ii)]
$\pi_2$ is positively oriented with respect to each of the lines
that support the edges of $\Delta_1$.
\end{itemize}

\medskip
Conditions (i) and (ii) are the conjunction of a total of six sub-conditions,
each of which tests the orientation of, say, the 2-plane $\pi_1$ with respect 
to the line supporting some specific edge of $\Delta_2$, or vice versa.

We can therefore apply a suitable variant of the same machinery of the 
preceding section, except that here all levels of the structure involves 
semi-algebraic range searching, with quadratic or cubic polynomial inequalities,
in six dimensions. This yields a proof of Theorem~\ref{thm:qsetup2}.

%---------------------------------------------
\paragraph{The batched bichromatic version.}
For the batched version of the triangle-triangle intersection problem, 
with $m$ red triangles and $n$ blue triangles (see Theorem~\ref{thm:setup2}), 
we choose the storage parameter $s$ to be such that the cost of $m$ queries with the red triangles is asymptotically 
roughly the same as the cost of preprocessing the blue triangles. That is, we set
$s = mn/s^{1/6}$, or $s = m^{6/7}n^{6/7}$. For this choice to make sense, we need 
to ensure that $n\le s\le n^6$, or that $n^{1/6} \le m\le n^6$. 
When $m > n^{6}$ we only use the data structure of \cite{AAEZ} and obtain 
the running time $O^*(m+n^6) = O^*(m)$, and when $m < n^{1/6}$ we only use 
the data structure of \cite{MP} and obtain the running time 
$O^*(mn^{5/6} + n) = O^*(n)$. Altogether we obtain the bound in 
Theorem~\ref{thm:setup2}.

% \micha{A grand challenge is to improve the algorithm for this setup too.}

%-----------------------------------------------------------------------------------
\section{Segment-Intersection amid Tetrahedra: \\ An Improved Solution}
\label{sec:shoot}

%\esther{This section should be shortened to about one page.}

In this section we present an improved algorithm for setup (i) of the paper, 
% where we improve the standard algorithm, presented in Section~\ref{app:trad},
for a data structure of roughly quadratic size.
Let $\T$ be a collection of $n$ tetrahedra in $\reals^4$.
Our improved solution constructs a data structure that uses $O^*(n^{2})$ 
storage (and expected preprocessing time), and answers a 
query in $O^*(n^{1/2})$ time. This is indeed a significant improvement over the 
standard algorithm in Section~\ref{app:trad}, in which, with storage $O^*(n^2)$, 
the query cost is $O^*(n^{2/3})$. With a suitable tradeoff, presented in 
Section~\ref{app:trade}, the improvement can be extended for any storage 
parameter between $n$ and $n^6$, although it is most substantial when the 
storage is nearly quadratic; see Figure~\ref{fig:trade}.

Assume, without loss of generality, that the query segment is bounded
(i.e., not a ray or a full line).
% Using similar notations as in Section~\ref{sec:bip},
The algorithm constructs a partitioning polynomial $F$ in $\reals^4$ of degree $O(D)$,
for some large but constant parameter $D$, so that each cell of the partition
is crossed by at most $n/D^2$ 2-faces of the tetrahedra in $\T$ and by a total of at most
$n/D$ tetrahedra. The existence of such a polynomial follows from Guth~\cite{Guth}, 
and an expected linear-time algorithm for its construction (for constant $D$) is given 
in \cite{AAEZ}. We classify each tetrahedron $\Delta\in\T$ as being \emph{narrow} 
(resp., \emph{wide}) with respect to a partition cell $\tau$ if a 2-face of 
$\Delta$ crosses $\tau$ (resp., $\Delta$ crosses $\tau$ but none of its 2-faces 
crosses $\tau$).
% \footnote{%
  % Note that this definition for the notions of wide and narrow tetrahedra
  % is somewhat different than the one for triangles as earlier defined in Section~\ref{sec:bip}.}
Let $\N_\tau$ (resp., $\W_\tau$) denote the set of narrow (resp., wide) tetrahedra at $\tau$.

% \esther{I rewrote this part onward to be suitable to the conference version.}

There are two cases to consider in our analysis, depending on whether the query 
segment $\rho$ is contained or not contained in the zero set $Z(F)$ of $F$.
Each of these cases requires its own data structure. The latter case 
is an extension of the analysis in~\cite{trishoot} (given there for the 
three-dimensional version of the problem), and the case where 
$\rho \subset Z(F)$ requires a different approach than that taken
in~\cite{trishoot} for handling queries on the zero set. See below for full details.

%--------------------------------------
\paragraph{A sketch of the analysis.}
A query segment $\rho$ that is not contained in $Z(F)$ crosses at most $O(D)$ 
cells of the partition. For each partition cell $\tau$ (an open connected 
component of $\reals^4\setminus Z(F)$) we construct an auxiliary data structure 
on the wide tetrahedra at $\tau$, and preprocess the narrow tetrahedra at
$\tau$ recursively. As we show in Section~\ref{sec:widex}, the structure for the wide tetrahedra 
uses $S_0(n) = O^*(n^2)$ storage, and a query amid them takes $Q_0(n) = O^*(n^{1/2})$ 
time. We then output some wide tetrahedron returned by querying the auxiliary 
structure at $\tau$, if such a tetrahedron exists. Otherwise, we return
the tetrahedron produced by the recursive call, if one exists. If no 
tetrahedron, wide or narrow, has been found, we proceed to the next cell\footnote{%
  The order of processing the cells is important for ray-shooting queries, but
  is immaterial for segment intersection queries.}
$\tau'$ crossed by $\rho$, repeat the whole procedure at $\tau'$, and keep
doing this till we either find a tetrahedron hit by $\rho$ or run out of cells,
and then conclude that $\rho$ does not hit any tetrahedron of $\T$.

The correctness of this procedure is clear (modulo that of the procedure
for handling wide tetrahedra). Denote by $S(n)$ (resp., $Q(n)$) the 
maximum storage (resp., query time) required by the overall structure for 
$n$ tetrahedra. Also denote by $S_1(n)$ (resp., $Q_1(n)$) the maximum 
storage (resp., query time) required for processing the input tetrahedra 
for intersection queries with segments contained in $Z(F)$, for any set 
of $n$ tetrahedra in $\reals^4$. We then have, for a suitable absolute 
constant $c > 0$ (where the constant hidden in the $O_D(\cdot)$ notation
depends on $D$),
\begin{align*}
  S(n) & = O_D(S_0(n/D)) + S_1(n) + c D^4 S(n/D^2) \\
  Q(n) & = \max\left\{ O_D(Q_0(n/D)) + cD Q(n/D^2) ,\; Q_1(n) \right\} .
\end{align*}
We show, in Section~\ref{sec:onzf}, that $S_1(n) = O_D^*(n^{2})$ and 
$Q_1(n) = O_D^*(n^{1/2})$. Substituting these bounds, as well as the 
bounds for $S_0(n)$ and $Q_0(n)$, the solutions of these recurrences 
are easily seen to be $S(n) = O^*(n^{2})$ and $Q(n) = O^*(n^{1/2})$.
Modulo the missing details, to be provided in what follows, this 
establishes Theorem~\ref{thm:main}.

%--------------------------------------------
\subsection{Handling the wide tetrahedra} \label{sec:widex}

Handling the wide tetrahedra resembles, and extends to four dimensions,
a similar machinery recently developed by the authors in \cite{trishoot}.
It is done via the following secondary recursion.
We choose some large constant parameter $r_0 \gg D$, and partition $\bd\tau$
into $O_D(1)$ $x_1x_2x_3$-monotone strata (assuming a generic choice of 
the coordinate frame). This is fairly standard to do, using the
\emph{cylindrical algebraic decomposition}~\cite{Col,SS2}, or CAD for 
short, of $F$ (see~\cite{BPR,Col,SS2} for details concerning this decomposition). 
We construct, for each stratum $\sigma$, a $(1/r_0)$-cutting for the set of 
(constant-degree algebraic) 2-surfaces of intersection of $\sigma$ with the 
wide tetrahedra in $\W_\tau$. The cutting is constructed by projecting 
$\sigma$ and the $2$-surfaces that it contains onto the $x_1x_2x_3$-subspace, 
constructing a $(1/r_0)$-cutting, within that subspace, on the projected 
surfaces, and then lifting the resulting cutting back to $\sigma$.
Using standard results on vertical decomposition in three dimensions (see, e.g.,
\cite{SA}) and the theory of cuttings~\cite{HW87}, we obtain $O^*(r_0^3)$ cells 
of the cutting (referred to as (pseudo-)prisms, in accordance with the way in 
which the vertical-decomposition--based cutting is constructed), each of which 
is crossed by (intersects but not contained in) at most $n/r_0$ wide tetrahedra.

For each pair $\psi_1$, $\psi_2$ of prisms, over all possible pairs of strata,
we define $S_{\psi_1,\psi_2}$ to be 
the set of all segments $e$ so that $e$ has an endpoint in $\psi_1$ and an endpoint 
in $\psi_2$, and the relative interior of $e$ is fully contained in $\tau$.
Clearly, $S_{\psi_1,\psi_2}$ is a semi-algebraic set of constant complexity
in a 6-dimensional parametric space,\footnote{%
  Each segment is specified by its two endpoints; since they lie on $\bd\tau$,
  each has three degrees of freedom.}
and we decompose it into its $O(1)$ connected components.

For each segment $e \in S_{\psi_1,\psi_2}$, let $\T(e)$ denote the set of all 
wide tetrahedra $\Delta$ of $\W_\tau$ that $e$ crosses. We have the following 
crucial technical lemma, akin to Lemma~2.2 in~\cite{trishoot}: 
%-----------------------------
\begin{lemma}
  \label{lem:cross4}
  Each connected component $C$ of
  $S_{\psi_1,\psi_2}$ can be associated with a fixed set $\T_C$ of
  wide tetrahedra $\Delta$ of $\W_\tau$, none of which 
  crosses $\psi_1\cup\psi_2$, so that, for each segment 
  $e\in C$, $\T_C\subseteq \T(e)$, and each tetrahedron $\Delta$
  in $\T(e)\setminus\T_C$ crosses either $\psi_1$ or $\psi_2$.
\end{lemma}
%-----------------------------

\begin{figure}[htb]
  \begin{center}
    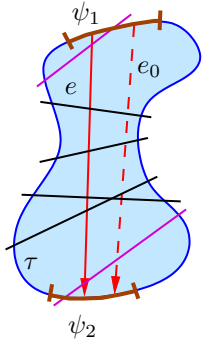
    \caption{{\sf The set $\T_C$ (consisting of the tetrahedra depicted as black segments),
        and an illustration of the proof of Lemma~\ref{lem:cross4}:
        The tetrahedra that cross some fixed segment $e_0$ between $\psi_1$ and $\psi_2$ are 
        the same tetrahedra that cross any other such segment $e$, except for those that cross 
        $\psi_1$ or $\psi_2$ (like those depicted as magenta segments).}}
    \label{fig:tcee0}
  \end{center}
\end{figure}

\noindent{\bf Proof.}
Pick an arbitrary but fixed segment $e_0$ in $C$, and define $\T_C$ to consist
of all the tetrahedra in $\T(e_0)$ that do not cross $\psi_1\cup\psi_2$.
See Figure~\ref{fig:tcee0} for an illustration.

Let $e$ be another segment in $C$. Since $C$ is connected, as a set in
the six-dimensional parametric space $\F$ of segments connecting a point
on $\psi_1$ with a point on $\psi_2$, there exists a continuous path $\pi$ 
in $C$ that connects $e_0$ and $e$. That is, each point on $\pi$ represents 
a segment with one endpoint on $\psi_1$ and the other on $\psi_2$, and $\pi$ 
represents a continuous variation of such a segment (in the Hausdorff metric 
sense) from $e_0$ to $e$.
 % \esther{what do we mean by that?} 
 % \micha{What does it mean that the segment moves continuously along the path?
  % That the mapping $t\mapsto e(t)$ that represents the path is continuous.
  % Under what metric? The Hausdorff is a natural measure.}
Let $\Delta$ be a tetrahedron in $\T(e_0)$ that does 
not cross $\psi_1\cup\psi_2$ (that is, $\Delta\in\T_C$). For a segment $e'\in \pi$, 
define the point $q_\Delta(e')$ to be the unique point $e'\cap \Delta$.
($q_\Delta(e')$ is indeed unique, if it exists, unless $e'$ gets to be 
contained in or partially overlap $\Delta$, a situation that we will shortly rule out.) 
As $e'$ starts traversing $\pi$ from $e_0$ to $e$, the point $q_\Delta(e')$ 
is well defined and varies continuously in $\tau\cap\Delta$, until we reach an 
instance at which either (i) the relative interior of $e'$ touches $\bd \Delta$, 
or (ii) an endpoint of $e'$ touches $\Delta$, or (iii) $e'$ comes to overlap 
$\Delta$ in an interval with a nonempty interior.

Case (i) cannot arise because the relative interior of $e'$ is fully contained in 
$\tau$ and $\Delta$ is wide at $\tau$. Case (ii) also cannot arise because then 
$\Delta$ would have to intersect either $\psi_1$ or $\psi_2$, which we have 
assumed not to be the case. Case (iii) is also impossible, because it implies that 
either Case (i) or Case (ii) must also arise, which cannot happen as just argued.

To recap, as $e'$ varies along $\pi$, it keeps intersecting $\Delta$ for every 
tetrahedron $\Delta\in \T_C$. Thus the endpoint $e$ of $\pi$ is also a segment
that crosses $\Delta$, and this establishes the first assertion of the lemma. 

We next need to show that each tetrahedron in $\T(e)\setminus\T_C$ must 
cross either $\psi_1$ or $\psi_2$, which is our second assertion.
Let $\Delta$ be a tetrahedron in $\T(e)\setminus\T_C$, and assume to the 
contrary that $\Delta$ does not cross $\psi_1\cup\psi_2$. We run the 
preceding argument in reverse (moving from $e$ to $e_0$), and observe 
that, by assumption and by the same argument (and notations) as above, 
$q_\Delta(e')$ remains well defined and inside $e'$, for all intermediate 
segments $e'$ along the connecting path $\pi$, and does not reach 
$\bd{(\Delta \cap \tau)}$, so $\Delta\in \T(e_0)$ and thus we have 
$\Delta\in \T_C$ (by definition of $\T_C$), contradicting our assumption. 
This establishes the second assertion, and thereby completes the proof.
$\Box$

%------------------------------------------------------
\paragraph{The analysis for wide tetrahedra.}

For each prism $\psi$, the \emph{conflict list} $K_\psi$ of $\psi$ is 
the set of all wide tetrahedra that cross $\psi$. By construction, 
$|K_\psi| \le n/r_0$. The same bound for crossing tetrahedra holds when
$\psi$ is lower-dimensional. If a lower-dimensional prism is contained in 
some tetrahedron there is no need to process $\psi$ further, since any segment 
that meets $\psi$ hits all these tetrahedra.

For each pair of prisms $\psi_1$, $\psi_2$, we compute $S_{\psi_1,\psi_2}$ 
and decompose it into its connected components. For each component $C$ we 
compute the set $\T_C$ of the wide tetrahedra, as in Lemma~\ref{lem:cross4}.
For this, we pick an arbitrary segment $e_0$ in $C$, compute the set $\T(e_0)$
as defined above, and remove from it all the tetrahedra that cross $\psi_1\cup\psi_2$.
All these operations can be implemented in $O_D(1)$, for a fixed pair
$\psi_1$, $\psi_2$, in the algebraic model that we assume (see~\cite{BPR}),
for a total of $O_D(r_0^6 n) = O_D(n)$ storage and computation time.

Let $s$ be the storage parameter associated with the problem; we require that $n\le s \le n^3$.
For each canonical set $\T_C$, we replace its tetrahedra by their supporting hyperplanes,
and preprocess the resulting collection of hyperplanes for efficient segment intersection 
queries amid hyperplanes in $\reals^4$.
Using the technique of Agarwal and Matou\v{s}ek~\cite{AM:ray},
this problem can be solved using $O^*(s)$ storage (and preprocessing), 
and a query takes $O(n\,\polylog(n)/s^{1/4}) = O^*(n/s^{1/4})$ time (see also~\cite{Ag:rs}).
Lemma~\ref{lem:cross4} guarantees the correctness of this procedure
(namely, that replacing each tetrahedron in $\T_C$ by its supporting hyperplane
does not cause any ``false positive'' answer).

We now process recursively each conflict list $K_\psi$, over all
prisms $\psi$ of the partition of $\bd\tau$. Each recursive subproblem uses
the same parameter $r_0$, but the allocated storage parameter is now $s/r_0^3$.
We keep recursing until we reach conflict lists of size close to $n^{3/2}/s^{1/2}$.
More precisely, after $j$ levels of recursion, we get a total of at most 
$(c_0 r_0^3)^j = c_0^jr_0^{3j}$ subproblems, each involving at most
$n/r_0^j$ wide tetrahedra, for some constant 
$c_0$ that depend on $D$ (but is considerably smaller than $r_0$,
which we ensure by taking $r_0$ to be sufficiently large).

We stop the recursion at the first level $j^*$ at which
$\frac{n}{r_0^{j^*}} \le n^{3/2}/s^{1/2}$.
As a result, we have ${r_0}^{j^{*}} = O(s^{1/2}/n^{1/2})$, and we get
$c_0^{j^{*}} r_0^{3j^{*}} = O^*(s^{3/2}/n^{3/2})$ subproblems.
Each of these subproblems involves at most 
$\frac{n}{r_0^{j^*}} = O^*\left( \frac{n^{3/2}}{s^{1/2}} \right)$
tetrahedra. Hence the overall size of the inputs, as well as of the canonical sets,
at all the subproblems throughout the recursion, is
$\displaystyle O^*\left(\frac{s^{3/2}}{n^{3/2}} \cdot \frac{n^{3/2}}{s^{1/2}} \right) = O^*(s)$.
In particular, this is the asymptotic cost at the bottom level of the recursion.

As just described, at the bottom of the recursion, each subproblem contains at most 
$O^*(n^{3/2}/s^{1/2})$ wide tetrahedra, and we detect intersections with them by brute force. 
We thus obtain the following recurrence for 
the overall storage $S_0(N_W,s_W)$ for the structure constructed on $N_W$ wide tetrahedra,
where $s_W$ denotes the storage parameter allocated to the structure
(at the root  $N_W = n$, $s_W = s$).
\[
S_0(N_W,s_W) = \left\{
\begin{array}{ll}
  O^*_D(r_0^6 s_W) +
    c_0r_0^3 S_0\left(\frac{N_W}{r_0},\; \frac{s_W}{r_0^3}\right) & \mbox{for $N_W \ge \Theta^*(n^{3/2}/s^{1/2})$,}\\[1mm]
  O(N_W) &\mbox{for $N_W < \Theta^*(n^{3/2}/s^{1/2})$}.
\end{array}
\right\}
\]
Unfolding the recurrence up to the terminal level $j^*$, where $N_W = O^*(n^{3/2}/s^{1/2})$, 
the sum of the nonrecursive overhead terms, over all nodes at a fixed level $j$, is
\[
c_0^{j} r_0^{3j} \cdot O^*\left( \frac{s_W}{r_0^{3j}} \right) =
O^*\left(s_W \right) .
\]
Hence, starting the recurrence at $(N_W,s_w) = (n,s)$, the overall contribution 
of the overhead terms is $O^*(s)$.
We showed above that this is also the asymptotic cost at the bottom of the recurrence.
Therefore, the overall storage used by the data structure is $O^*(s)$.
Using similar considerations, one can show that the overall expected preprocessing 
time is $O^*(s)$ as well, since the time obeys a similar asymptotic recurrence.

%------------------------------
\paragraph{Answering a query.}

Given a query segment $\rho$, which is not contained in $Z(F)$, we find its $O(D)$ 
intersections with $Z(F)$, which decompose it into $O(D)$ segments, each fully
contained in some partition cell. Moreover, except for the first and last segment,
the endpoints of each of the other segments lie on the boundary of its cell. 
We process the segments in their order\footnote{%
  As already mentioned, the order is immaterial for segment intersection 
  detection queries, but is important for ray shooting.}
along $\rho$. Let $e$ be the currently
processed segment. If $e$ is not the first or last segment, we find the prisms 
$\psi_1$, $\psi_2$ that contain its endpoints, and find the component $C$ of 
$S_{\psi_1,\psi_2}$ that contains $e$. If $e$ is the first or last segment, we extend 
it backwards or forwards, respectively, till the first time it meets the boundary of 
its cell, and call the resulting segment $e'$.
We now compute for $e'$ the corresponding set $S_{\psi_1,\psi_2}$ and its component 
$C$ that contains $e'$. Since $D$ and $r_0$ are constants, all this takes constant time.

The query, on the wide tetrahedra at the present cell $\tau$, performs a segment
intersection detection query with $e$ (or with $e'$ when $e$ is the first or last
segment) in the set of hyperplanes containing the tetrahedra of $\T_C$, and, if 
no intersection is detected, continues recursively with the conflict lists
$K_{\psi_1}$ and $K_{\psi_2}$ (at the bottom of recursion we apply a brute-force search). 
If no tetrahedron is found, in all the $r_0$-recursive steps, we
conclude that (the present subsegment of) $\rho$ does not hit any wide
tetrahedron within $\tau$. Once again, the correctness of this procedure 
follows from Lemma~\ref{lem:cross4}.

The query time $Q_0(N_W,s_W)$ satisfies the recurrence
\[
Q_0(N_W,s_W) = \left\{
\begin{array}{ll}
  O_D(1) + O^*\left(\frac{N_W} {s_W^{1/4}} \right) 
+ 2Q\left(\frac{N_W}{r_0},\, \frac{s_W}{r_0^3} \right) & \mbox{for $N_W \ge \Theta^*(n^{3/2}/s^{1/2})$,}\\[1mm]
  O(N_W) & \mbox{for $N_W < \Theta^*(n^{3/2}/s^{1/2})$} .
\end{array}
\right\}
\]
Unfolding the recurrence, the overall bound for the nonrecursive overhead terms,
starting from $(N_W,s_W) = (n,s)$, is at most
\[
O^*\left( \sum_{j\ge 0}
\left( \frac{2}{r_0^{1/4}} \right)^{j} \cdot \frac{n}{s^{1/4}} \right) =
O^*\left( \frac{n}{s^{1/4}} \right) .
\]
Adding the cost at the $2^{j^*}$ subproblems
at the bottom level $j^*$ of the recursion, where the cost of each subproblem is at 
most $O^*(n^{3/2}/s^{1/2})$, we obtain the query time 
\begin{equation}
  \label{eq:qns}
  Q_0(n,s) = O^*\left( \frac{n}{s^{1/4}} + \frac{n^{3/2}}{s^{1/2}} \right) .
\end{equation}
Therefore, for $s=n^{2}$ the query time is $O^*(n^{1/2})$.
The bounds $S_0(n) := S_0(n, n^2) = O^*(n^{2})$ and $Q_0(n) := Q_0(n, n^2) = O^*(n^{1/2})$
are the bounds promised earlier for the wide tetrahedra at a cell.

%----------------------------------
\subsection{Query segments on $Z(F)$} \label{sec:onzf}

Consider next segments $\rho$ that are contained in $Z(F)$. Without loss of
generality we may assume that $F$ is irreducible; otherwise we apply the
forthcoming machinery separately to each irreducible factor of $F$.
(Decomposing $F$ into its irreducible factors, over the reals, can be done
in $O_D(1)$ time in the algebraic model that we are using.)
% \micha{Pankaj was fighting with this issue. Check and cite something.}
We may also assume that $Z(F)$ is not a hyperplane. If it is, we simply
face an intersection detection problem in three dimensions amid a collection
of triangles (each tetrahedron crosses $Z(F)$ in a convex polytope of 
constant complexity, and we replace it by its triangulated boundary).
This latter task has been studied in \cite{trishoot}, where a solution 
with better performance bounds has been given.

We partition $Z(F)$ into $O_D(1)$ $x_1x_2x_3$-monotone strata, as we
did in the algorithm for wide tetrahedra. These strata covers $Z(F)$ 
for a generic choice of the coordinate frame. Each tetrahedron 
$\Delta\in\T$ intersects $Z(F)$ in a semi-algebraic set $\Delta_F$ 
of constant complexity (that depends on $D$), and we distribute it among 
all the strata that it intersects. We project each stratum $\sigma$, and the 
portions of the sets $\Delta_F$ that it contains, onto the $x_1x_2x_3$-space. 
For a stratum $\sigma$ and a tetrahedron $\Delta$ that crosses $\sigma$, we denote 
the $x_1x_2x_3$-projection of $\Delta\cap\sigma$ (i.e., $\Delta_F\cap\sigma$),
which is also a semi-algebraic set of constant complexity, as $K_\Delta$.
We also denote by $B_\Delta$ the $x_1x_2x_3$-projection of the intersection 
of $\bd\Delta$ with $\sigma$; note that $B_\Delta$ is the union of up to four 
subsets, each of which is the intersection of a different 2-face of $\Delta$
with $Z(F)$. % \micha{Handle tetrahedra that are fully contained in $Z(F)$?}
Excluding degenerate scenarios, each $K_\Delta$ is two-dimensional, 
and each $B_\Delta$ is one-dimensional. Indeed, $\Delta\cap Z(F)$ is two-dimensional 
(unless $\Delta$ is fully contained in $Z(F)$), and each of the four subsets of 
$B_\Delta$ is contained in the intersection of a 2-plane with $Z(F)$, which is a 
constant-degree algebraic curve (unless this 2-plane is fully contained in $Z(F)$).
Handling the degenerate cases, where $\Delta$ or one of its 2-faces is fully 
contained in $Z(F)$, is easier, and is omitted here. % \micha{Check that theree is no issue here.}

We thus face the problem of segment intersection detection in three 
dimensions (the query segment projects to a segment in 3-space) amid 
a collection $\K$ of $n$ two-dimensional semi-algebraic sets of constant complexity. 
We are not aware of an efficient solution to this problem. (A standard 
solution that maps the problem to semi-algebraic range searching in a 
higher-dimensional parametric space, results in a much less efficient solution.)
We obtain a fast procedure by exploiting several special properties of our
setting. Specifically, we exploit two constraints on the problem: 

\medskip
\noindent
(a) The sets in $\K$ have a special structure---each of them is the 
$x_1x_2x_3$-projection of the intersection of a tetrahedron with $Z(F)$. 

\medskip
\noindent
(b) The query segments also have a special structure---each such segment
is the $x_1x_2x_3$-projection of a segment supported by a line that is fully
contained in $Z(F)$. 

\medskip
To exploit property (b), let $\L$ denote the
set of all lines that are fully contained in $Z(F)$, and let $\L^*$
denote the set of the $x_1x_2x_3$-projections of these lines. Since
$F$ is of constant degree, standard arguments in real algebraic geometry
imply that $\L^*$ is a semi-algebraic set of constant complexity;
see, e.g., \cite{BPR}.

We tackle this problem by processing each stratum $\sigma$ in turn.
We construct a (second) trivariate partitioning polynomial $G$, of 
degree $O(D_1)$, where $D_1\gg D$ is another constant parameter, so that each 
cell of $\reals^3\setminus Z(G)$ is crossed by at most $n/D_1^2$ one-dimensional 
curves $B_\Delta$, and by at most $n/D_1$ two-dimensional sets $K_\Delta$.

A query segment $\rho$ contained in $Z(F)$ is projected to a segment $\rho^*$ 
in $\reals^3$ (the $x_1x_2x_3$-space), whose supporting line belongs
to $\L^*$. Two cases can arise:

%-----------------------------------------------
\paragraph{$\rho^*$ is not contained in $Z(G)$.}
We say that a tetrahedron $\Delta\in\T$ is \emph{narrow} at a cell $\tau$ 
of the partition induced by $G$ if $B_\Delta$ crosses $\tau$, and $\Delta$ 
is \emph{wide} at $\tau$ if $K_\Delta$ crosses $\tau$ but $B_\Delta$ does not. 
As in the four-dimensional case, we denote by $\W_\tau$ (resp., by $\N_\tau$) 
the set of wide (resp., narrow) tetrahedra at $\tau$. We preprocess the wide 
tetrahedra at $\tau$ using a special substructure, and handle the narrow 
tetrahedra recursively.

%----------------------------------------
\paragraph*{Handling the wide tetrahedra.}
The analysis is similar to that for wide tetrahedra in four dimensions,
as presented in Section~\ref{sec:widex}, but we spell it in detail, risking
repetition of some of the arguments, as the actual technical details
are different in the current setup.

Let $\tau$ be a cell of $\reals^3\setminus Z(G)$. We partition $\bd\tau$ 
into $O^*(r_0^2)$ pseudo-trapezoids (trapezoids for short), 
for some suitable constant parameter $r_0 \gg E$, so that each trapezoid
is crossed by at most $|\W_\tau|/r_0$ tetrahedra of $\W_\tau$.
For each pair $\psi_1$, $\psi_2$ of trapezoids, we define 
$S_{\psi_1,\psi_2}$ to be the set of all segments $e$ so that 
(a) $e$ has an endpoint in $\psi_1$ and an endpoint in $\psi_2$,
(b) the relative interior of $e$ is fully contained in $\tau$, and
(c) the line supporting $e$ belongs to $\L^*$. Clearly, $S_{\psi_1,\psi_2}$ 
is a semi-algebraic set of constant complexity, as each of the conditions 
(a)--(c) can be expressed as a semi-algebraic predicate of constant
complexity, possibly using quantifiers. We decompose $S_{\psi_1,\psi_2}$ 
into its $O(1)$ connected components.

For each segment $e \in S_{\psi_1,\psi_2}$, let $\T(e)$ denote the
set of all wide tetrahedra $\Delta$ of $\W_\tau$ such that $e$ crosses
their associated sets $K_\Delta$. As in the four-dimensional case,
our technique depends on the following crucial technical lemma.
(Intuitively, the intersections of the wide tetrahedra with $Z(f)$ 
have the crucial property, which is needed in the proof, that any 
segment on $Z(f)$ meets each of them only once. This does not necessarily 
hold in the 3D projection, but, as we argue below, this does not hurt the analysis.
%---------------------
\begin{lemma}
  \label{lem:cross3}
  Each connected component $C$ of
  $S_{\psi_1,\psi_2}$ can be associated with a fixed set $\T_C$ of
  wide tetrahedra $\Delta$ of $\W_\tau$, none of whose associated sets 
  $K_\Delta$ crosses $\psi_1\cup\psi_2$, so that, for each segment 
  $e\in C$, $\T_C\subseteq \T(e)$, and for each tetrahedron $\Delta$
  in $\T(e)\setminus\T_C$, $K_\Delta$ crosses either $\psi_1$ or $\psi_2$.
\end{lemma}
%---------------------

\begin{figure}[htb]
  \begin{center}
    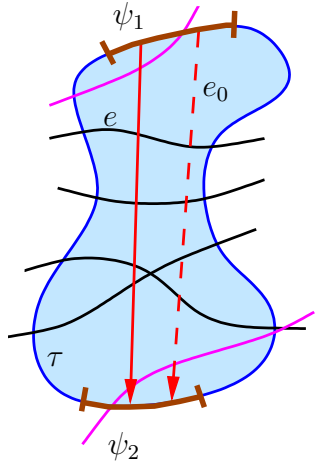
    \caption{{\sf The set $\T_C$ (consisting of the tetrahedra $\Delta$ whose 
    associated sets $K_\Delta$ are depicted as black arcs),
    and an illustration of the proof of Lemma~\ref{lem:cross3}.}}
    \label{fig:tcee1}
  \end{center}
\end{figure}

\noindent{\bf Proof.}
Pick an arbitrary but fixed segment $e_0$ in $C$, and define $\T_C$ to consist
of all the tetrahedra in $\T(e_0)$ that do not cross $\psi_1\cup\psi_2$.
See Figure~\ref{fig:tcee1} for an illustration.

Let $e$ be another segment in $C$. The set $S_{\psi_1,\psi_2}$ has four
degrees of freedom, two for representing the endpoint of a segment $e$ 
in $S_{\psi_1,\psi_2}$ that lies on $\psi_1$, and two for the other
endpoint (on $\psi_2$). Since $C$ is connected, as a subset of
$S_{\psi_1,\psi_2}$, there exists a continuous path $\pi$ in $C$ 
that connects $e_0$ and $e$. (As in the four-dimensional case, each 
point on $\pi$ represents a segment with one endpoint on $\psi_1$ and 
the other on $\psi_2$, which is contained in a line of $\L^*$ and is 
fully contained in the interior of $\tau$, and $\pi$ represents a 
continuous variation of such a segment from $e_0$ to $e$.) 

Let $\Delta$ be a tetrahedron in $\T(e_0)$ such that $K_\Delta$ 
does not cross $\psi_1\cup\psi_2$ (that is, $\Delta\in\T_C$). 
% \micha{The previous argument here was wrong, because $e'$ might intersect
% $K_\Delta$ at more than one point---the other points are intersections 
% of $e'$ with $K_\Delta$ that do not come from intersections in $\reals^4$.
% I tried to fix. Please check very carefully.} 
For $e_0$, the intersection $e_0\cap K_\Delta$, if nonempty, contains 
the projection of the single intersection point of the pre-image of $e_0$ 
with $\Delta$ (and might also contain additional points).
We denote this point as $q_\Delta(e_0)$. As $e'$ varies along $\pi$ from
$e_0$ towards $e$, the corresponding point $q_\Delta(e')$ is well defined
and varies continuously in $\tau$, until we reach an instance at which either 
(i) the relative interior of $e'$ touches $\bd K_\Delta$, or (ii) $e'$ 
becomes tangent to $K_\Delta$, or (iii) an endpoint of $e'$ touches 
$K_\Delta$, or (iv) $e'$ comes to overlap $K_\Delta$ in an interval 
with a nonempty interior.

Case (i) cannot arise because the relative interior of $e'$ is fully contained in 
$\tau$ and $K_\Delta$ is wide at $\tau$. Case (iii) cannot arise because then 
$K_\Delta$ would have to intersect either $\psi_1$ or $\psi_2$, which we have 
assumed not to be the case. Case (ii) can occur, but then, assuming that cases 
(i) and (iii) do not occur at the same time, the line $\ell$ in $\reals^4$, 
which is contained in $Z(F)$ and projects to the line supporting $e'$, continues
to cross $\Delta$, and its intersection point with $\Delta$ continues to project
to a point in $K_\Delta$ (because $\ell\subset Z(F)$). That is, as we continue to 
vary $e'$ further towards $e$, the line $\ell$ changes continuously and keeps 
crossing $\Delta$ (because Case (iii) does not arise), and thus $e'$ also keeps 
crossing $K_\Delta$ (because Case (i) does not arise). That is, the instantaneous
tangency does not cause $q_\Delta(e')$ to disappear, or to experience any jump 
discontinuity. In an instance of Case (iv), which is not an instance of Case (i),
Case (ii), or Case (iii), the line $\ell$ must be fully contained in the hyperplane
supporting $\Delta$, so the projection of $\ell$ cuts $K_\Delta$ in a connected
segment, from which it easily follows that one of Cases (i), (iii) must arise 
for $e'$, a contradiction that takes care of this case too.

To recap, as $e'$ varies along $\pi$, it keeps intersecting $K_\Delta$ for every 
tetrahedron $\Delta\in \T_C$. Thus the endpoint $e$ of $\pi$ is also a segment
that crosses $K_\Delta$, and this establishes the first assertion of the lemma. 

We next need to show that, for each tetrahedron $\Delta\in\T(e)\setminus\T_C$,
the associated set $K_\Delta$ must cross either $\psi_1$ or $\psi_2$, which is 
our second assertion. Let $\Delta$ be a tetrahedron in $\T(e)\setminus\T_C$, 
and assume to the contrary that $K_\Delta$ does not cross $\psi_1\cup\psi_2$. 
We run the preceding argument in reverse (moving from $e$ to $e_0$), and observe 
that, by assumption and by the same argument (and notations) as above, 
$q_\Delta(e')$ remains inside $e'$, for all intermediate segments $e'$ 
along the connecting path $\pi$, and does not reach $\bd{K_\Delta \cap \tau}$,
so $\Delta\in \T(e_0)$ and thus we have $\Delta\in \T_C$ (by definition
of $\T_C$), contradicting our assumption. 
This establishes the second assertion, and thereby completes the proof.
$\Box$

\medskip

For each pair of trapezoids $\psi_1$, $\psi_2$, and each connected component 
$C$ of $S_{\psi_1,\psi_2}$, we take the set $\T_C$ of tetrahedra (back 
in $\reals^4$), replace each $\Delta\in\T_C$ by its supporting
hyperplane, and preprocess the resulting collection of hyperplanes
for efficient segment intersection detection amid hyperplanes in $\reals^4$. 
Using the technique of \cite{AM:ray},
% \esther{I erased the text exaplaining the reduction to dihedral wedge emptiness, and just cite~\cite{AM:ray}.}
% this becomes an emptiness range searching problem with a dihedral wedge in a set 
% of $n$ points in $\reals^4$. As in the four-dimensional setup (when $\rho$ is not contained in $Z(F)$), with 
this can be done, with $O^*(s)$ storage and preprocessing, with
query time $O^*(n/s^{1/4})$. Choosing $s = n^2$, the storage complexity 
is $O^*(n^{2})$ and the query time is $O^*(n^{1/2})$.
Lemma~\ref{lem:cross3} guarantees the correctness of this procedure
(namely, of replacing each tetrahedron by its supporting hyperplane).

We then preprocess recursively each of the sets $\T_\psi$, of the
tetrahedra $\Delta$ for which $K_\Delta$ crosses $\psi$, over all
trapezoids $\psi$ of the partition of $\bd\tau$. A query, with a segment
$\rho$ that is contained in $Z(F)$ but its projection $\rho^*$ is not
contained in $Z(G)$, is then processed as follows. As in the 
four-dimensional setup, we need a special treatment for the first and last
subsegments of $\rho$, but we omit here the straightforward details, which
are similar to those in the preceding analysis.
We first perform a segment intersection detection query in the set 
of hyperplanes of the tetrahedra in $\T_C$, for the suitable component 
$C$ that contains the intersection segment of $\rho^*$ and $\tau$, 
and then continue recursively with $\T_{\psi_1}$ and $\T_{\psi_2}$, 
where $\psi_1$ and $\psi_2$ are the trapezoids that contain the 
endpoints of the segment. We stop the recursion at nodes $\psi$ for 
which $|\T_\psi|$ becomes roughly $n^2/s$. If no intersection with any
wide tetrahedron has been detected, we query recursively the set of 
narrow tetrahedra at $\tau$. If no tetrahedron is found to intersect 
the present portion of $\rho$ within $\tau$, we proceed to the next 
cell $\tau'$ crossed by the projected segment, and keep doing this until 
we either find a tetrahedron intersected by $\rho$, or run out of cells, 
and then conclude that $\rho$ does not intersect any tetrahedron of $\T$.

The correctness of this procedure is clear. The storage and the 
query cost obey two respective recurrences, one from the recursion
on wide tetrahedra at a cell, and one from the recursion on narrow
tetrahedra. For the first recurrence, denote by $S'_0(N_W,s_W)$ the 
maximum storage required by the structure for $N_W$ wide tetrahedra, 
where $s_W$ is the storage parameter allocated to the structure. We then have 
\[
S'_0(N_W,s_W) = \left\{
\begin{array}{ll}
  O^*_{D_1}(r_0^4 s_W) +
  c_0 r_0^2 S'_0\left(\frac{N_W}{r_0},\; \frac{s_W}{r_0^2}\right) & \mbox{for $N_W \ge \Theta^*(n^{2}/s$),}\\[1mm]
  O(N_W) &\mbox{for $N_W < \Theta^*(n^{2}/s)$}.
\end{array}
\right\}
\]
The constant $c_0$ depends on ${D_1}$, but is considerably smaller than $r_0$
(that is, we choose $r_0$ to be considerably larger).
We also comment that throughout this recursion $N_W \le s_W \le N_W^2$.
The terminal level $j^*$ of the recurrence
satisfies ${r_0}^{j^{*}} \le s/n$. It is then easily checked that 
the total contribution of all the overhead terms,
as well as the terms at the bottom of the recurrence, is $O^*(s)$, 
where $n \le s \le n^2$. Therefore the overall storage used by the data structure is $O^*(s)$.

Concerning the query time, denote by $Q'_0(N_W,s_W)$ the maximum query 
time required by the structure for $N_W$ wide tetrahedra. We then have:
\[
Q'_0(N_W,s_W) = \left\{
\begin{array}{ll}
  O_{D_1}(1) + O^*\left(\frac{N_W} {s_W^{1/4}} \right) 
+ 2Q\left(\frac{N_W}{r_0},\; \frac{s_W}{r_0^2} \right) & \mbox{for $N_W \ge \Theta^*(n^{2}/s$),}\\[1mm]
  O(N_W) & \mbox{for $N_W < \Theta^*(n^{2}/s)$} .
\end{array}
\right\}
\]
Unfolding the recurrence, the overall contribution of the non-recursive 
overhead terms (up to the bottom level $j^*$) is:
\[
O^*\left( \sum_{j\ge 0}
\left( \frac{2}{r_0^{1/2}} \right)^{j} \cdot \frac{n}{s^{1/4}} \right)  =
O^*\left( \frac{n}{s^{1/4}} \right) .
\]
The cost at the bottom level $j^*$ of the recursion is at most $O^*(n^{2}/s)$ (by the choice of $j^*$).
This yields an overall bound for the query time of
\begin{equation}
  \label{eq:qns_0}
  Q'_0(n,s) = O^*\left( \frac{n}{s^{1/4}} + \frac{n^{2}}{s} \right) .
\end{equation}
We thus obtain $S'_0(n) := S_0(n, n^2) = O^*(n^{2})$ and $Q'_0(n) := Q_0(n, n^2) = O^*(n^{1/2})$
for the overall storage and query cost of this subprocedure.

For the second recurrence, on narrow tetrahedra, denote, as above, by $S_1(n)$ 
(resp., $Q_1(n)$) the maximum storage (resp., query time) required by the 
structure for $n$ tetrahedra. We then have 
\begin{align*}
  S_1(n) & = O_{D_1}(S'_0(n/{D_1})) + S_2(n) + O({D_1}^3) S_1(n/{D_1}^2) \\
  Q_1(n) & = \max \Bigl\{ O_{D_1}(Q'_0(n/{D_1})) + ({D_1}+1) Q_1(n/{D_1}^2) ,\; Q_2(n) \Bigr\} ,
\end{align*}
where $S_2(n)$ (resp., $Q_2(n)$) is the maximum storage (resp., query time)
for segments $\rho$ such that $\rho\subset Z(F)$ and $\rho^*\subset Z(G)$.
As we show next, these quantities satisfy the bounds $S_2(n) = O^*(n^{2})$
and $Q_2(n) = O^*(n^{1/2})$. With these bounds at hand, the solutions of
these recurrences are $S_1(n) = O^*(n^{2})$ and $Q_1(n) = O^*(n^{1/2})$.

%-------------------------------------------
\paragraph{$\rho^*$ is contained in $Z(G)$.}

It remains to handle query segments $\rho^*$ that are contained in $Z(G)$.
We may assume that $Z(G)$ is irreducible; otherwise we apply the
following reasoning within each irreducible component of $Z(G)$.
As $Z(G)$ is a two-dimensional algebraic surface of degree $O({D_1})$, 
it is either ruled (by lines) or not ruled. In the latter case, $Z(G)$
contains only $O({D_1}^2)$ lines, as implied by the Cayley-Salmon theorem~\cite{salmon},
and we can prepare the answers to all possible queries along such lines
(only for those lines that belong to $\L^*$). Although not a trivial
step, all these lines can be computed in $O_{D_1}(1)$ time, which is constant
since ${D_1}$ is constant, by solving a suitable set of equations that
characterize these lines; see \cite{salmon} and \cite{BPR}.

In the former case, $Z(G)$ is either singly ruled, or doubly ruled (a regulus),
or infinitely ruled (a plane).
Assume first that $Z(G)$ is singly ruled. Then (see, e.g., \cite{GK}), except 
for at most two exceptional lines, the lines ruling $Z(G)$ form a 1-parameter 
family of lines. For each tetrahedron $\Delta$, the set of parameters of the 
lines whose pre-images, back in $\reals^4$, cross $\Delta$ is the union of
$O_{D_1}(1)$ intervals, as is easily verified, and they can all be computed in
$O_{D_1}(1)$ time. 
%\esther{We need to say how do we compute these intervals.}
We store the $O_{D_1}(n)$ resulting intervals, obtained over all tetrahedra 
$\Delta\in\T$, in a segment tree. For each node $\nu$ of the tree, 
we take the set $\T_\nu$ of tetrahedra stored at $\nu$ and preprocess 
the set $H_\nu$ of the hyperplanes supporting these tetrahedra into a 
segment intersection data structure based on the machinery in~\cite{AM:ray},
as in the previous steps of the algorithm.
% \esther{I revised the following part in order to obtain storage/query bounds that depend on $n$ and $s$.}
We allocate a storage parameter $s$ to each level of the segment tree just constructed.
At each node $\nu$, at any fixed level of the tree, we allocate a storage parameter
that is proportional to the number of tetrahedra stored at $\nu$. Specifically, 
we allocate $s \cdot \frac{|H_\nu|}{n}$ to $\nu$. In this manner we obtain a 
segment-intersection data structure at $\nu$ that uses $O^*(s |H_\nu|/n)$ storage and 
answers a query in time
\[
O^*\left( \frac{|H_\nu|}{\left(s \cdot \frac{|H_\nu|}{n}\right)^{1/4}} \right) =
O^*\left( \frac{|H_\nu|^{3/4} n^{1/4}}{s^{1/4}} \right) =
O^*\left(\frac{n}{s^{1/4}} \right) ,
\]
since $|H_\nu| \le n$.
A query with a segment $\rho$ finds the atomic (leaf) interval of the tree that 
contains the line supporting $\rho$, retrieves the $O(\log n)$ nodes on the path to
that leaf, performs segment intersection queries with $\rho$ in the sets
$H_\nu$ of these nodes $\nu$, and returns a tetrahedron from the output to these
queries (if such a tetrahedron exists).

It is easily checked, using standard properties of segment trees (i.e., that 
each tetrahedron appears in at most two sets $\T_\nu$ at any level of 
the tree), that the overall storage used by this structure is $O^*(s)$,
and that the query cost is $O^*(n/s^{1/4})$.

The case where $Z(G)$ is doubly ruled (a regulus) is handled similarly, applying 
the above machinery to each of the two families of ruling lines, each of which 
has a very simple structure.

The case where $Z(G)$ is a plane is easier to handle. In this case we do not
lift the scenario back to $\reals^4$ but instead remain on the plane $Z(G)$,
and face there the problem of segment intersection detection amid a collection 
of $n$ constant-degree algebraic arcs in the plane, where the arcs are the
intersections of the sets $K_\Delta$ with $Z(G)$. This can be done with 
$O^*(s)$ storage (and preprocessing) and $O^*(n^{2}/s)$ query time.\footnote{%
  In~\cite{trishoot}, the storage and query bounds 
  are $O^*(n^{3/2})$ and $O^*(n^{1/2})$, respectively, but a closer inspection 
  at the analysis shows that one can get the more general bounds stated above.}
We thus achieve in this case faster query time,
since $n^{2}/s \le n^{3/2}/s^{1/2}$, when $n \le s \le n^3$.

% \micha{Why didn't we use the same reasoning in the trishoot paper?! 
% Need to discuss this, and, if I am right, to revise the corresponding part of that paper considerably.}
% \esther{At that time we didn't think about it, but, on the other hand, the machinery for the zero set
  % in \cite{trishoot} is quite general and works for any collection of semi-algebraic sets in the plane,
  % I would keep it.}

Summarizing all the above cases, we indeed obtain that for $s = n^2$, the resulting
storage and query time bounds are $S_2(n) = O^*(n^{2})$ and $Q_2(n) = O^*(n^{1/2})$.
As already noted, this implies that the solutions of the preceding recurrences for 
$S_1(n)$ and $Q_1(n)$ are $S_1(n) = O^*(n^{2})$ and $Q_1(n) = O^*(n^{1/2})$.
We thus have finally completed the proof of Theorem~\ref{thm:main}.

\medskip
\noindent{\bf Remark.}
Informally, the reason why we have managed to improve the solution only for Setup (i) 
(for segment queries) is that when the queries are triangles (in Setup (ii)) or tetrahedra 
(in Setup (iii)), the query object intersects too many cells of the polynomial partition,
and the resulting recurrence for the query time does not yield any more efficient solution.
It is an interesting open challenge to find improved solutions for these setups too.

%-------------------------------------------------
\section{Tradeoff between Storage and Query Time}
\label{app:trade}

In this section we extend the technique in Section~\ref{sec:shoot} to
obtain a tradeoff between storage (and expected preprocessing) and query time,
which improves the standard tradeoff of Theorem~\ref{thm:standard}, for any value
$n\le s\le n^6$ of the storage parameter.

For a quick overview of our approach, consider the segment-intersection structure of 
Section~\ref{sec:shoot}, and let $s$ be the storage parameter that we allocate 
to the structure, which now satisfies $n\le s\le n^{6}$. We modify the procedure 
for segment intersection inside a cell $\tau$ by (i) stopping potentially the $r_0$-recursion 
at some earlier `premature' level, and (ii) modifying the structure at the bottom 
of recursion so that it uses the segment-intersection technique for hyperplanes, as discussed
in Section~\ref{app:trad}, instead of a brute-force scanning of the tetrahedra 
(the current cost of $O(n^{3/2}/s^{1/2})$, a consequence of this brute-force approach,
is too expensive when $s$ is small). A similar adaptation is applied to the 
recursion on the narrow triangles, as well as the procedure of segment intersection within
the zero set of the partitioning polynomial. 
With some additional care we obtain the query time bound in (\ref{eq:trade1}) 
and the bound (\ref{eq:trade2}) for batched segment intersection, as announced in the introduction.

We now present the technique in detail.
Consider the segment-intersection structure of Section~\ref{sec:shoot}, and let 
$s$ be the storage parameter that we allocate to the structure, which
satisfies $n\le s\le n^{6}$. As before, we use this notation to indicate 
that the actual storage (and expected preprocessing) that the structure uses may 
be $O^*(s)$. We comment that in the preceding Section~\ref{sec:shoot} $s$ is assumed 
to be (at most) $n^2$. Handling larger values of $s$ require some care, 
detailed below. For the time being, we continue to assume that $s\le n^2$, 
and will later show how to extend the analysis for larger values.

Consider first the subprocedure for handling segment intersection for segments that
are not contained in the zero set of the partitioning polynomial.
% 
% When $s = O(n^{3})$, 
We run the recursive polynomial partitioning procedure 
described in Section~\ref{sec:shoot} up to some `premature' level $k$ that 
we will fix later.
% and when $s = \Omega(n^{3})$, we run it all the way down. 
% In the former situation, 
We obtain $O^*(D^{4k})$ subproblems at the bottom 
level of recursion, each involving at most $n/D^{2k}$ (narrow) tetrahedra. 

%------------------------------------
\paragraph{Handling wide tetrahedra.}
Except for the bottom level, 
we build, at each node $\tau$ of the recursion, the same structure on 
the set $\W_\tau$ of wide tetrahedra in $\tau$, with two (significant) 
differences. First, since we start the recursion on the partitioning with 
storage parameter $s$, we allocate to each subproblem, at any level $j$, 
the storage parameter $s/D^{4j}$, thus ensuring that the storage used by the 
structure is $O^*(s)$. However, the cost of a query, even at the first 
level of recursion, given in (\ref{eq:qns}), has the term $O^*(n^{3/2}/s^{1/2})$,
which is the cost of a na\"ive, brute-force processing of the conflict lists 
at the bottom instances of the $r_0$-recursion within the partition cells.
This is fine for $s = \Omega^*(n^{2})$ but kills the efficiency of the
procedure when $s$ is smaller. For example, for $s=n$ we get (near) linear 
query time, much more than what we aim to have. We therefore improve the 
performance at the bottom-level nodes of the $r_0$-recurrence (within a 
partition cell), by constructing, for each respective conflict list, the 
segment-intersection data structure of Section~\ref{app:trad} for segment intersection
amid hyperplanes in $\reals^4$, which, for $N$ tetrahedra and with storage parameter 
$s$, answers a query in time $O^*(N/s^{1/6})$. Since at the bottom of the 
$r_0$-recursion, both the number of tetrahedra and the storage parameter are 
$O^*(n^{3/2}/s^{1/2})$, the cost of a query at the bottom of the recursion is 
\[
O^*((n^{3/2}/s^{1/2})^{5/6}) = O^*(n^{5/4}/s^{5/12}) .
\]
That is, the modified (improved) cost of a query at such a node is
\begin{equation}
  \label{eq:qns1}
  Q(n,s) = O^*\left( \frac{n}{s^{1/4}} + \frac{n^{5/4}}{s^{5/12}} \right) .
\end{equation}

% \micha{I went over this, carefully, until this point. The rest of the tradeoff is buggy, 
  % with wrong exponents. Please take over and shaptzi.}
% \esther{I fixed the bounds before this point and after.}

\paragraph{Handling the recursion on the polynomial partitions.}
At each of the $O^*(D^{4k})$ bottom-level cells $\tau$, we take the 
set $\N_\tau$ of (narrow) tetrahedra that have reached $\tau$, whose size is now 
at most $n/D^{2k}$, allocate to it the storage parameter $s/D^{4k}$, 
and preprocess $\N_\tau$ using the aforementioned technique of Section~\ref{app:trad},
which results in a data structure, with storage 
parameter $s/D^{4k}$, which supports segment-intersection queries in time 
\[
O^*\left(\frac{|\N_\tau|}{(s/D^{4k})^{1/6}}\right) = 
O^*\left(\frac{n/D^{2k}}{(s/D^{4k})^{1/6}}\right) = 
O^*\left(\frac{n}{s^{1/6}D^{4k/3}} \right) .
\]
Multiplying this bound by the number $O^*(D^{k})$ of cells that the
query segment crosses, the cost of the query at the bottom-level cells is 
\[
Q_{\rm bot}(n,s) = O^*\left(\frac{n}{s^{1/6}D^{k/3}} \right) .
\]
The cost of a query at the inner recursive nodes of some depth $j<k$
is the number, $O^*(D^{j})$, of $j$-level cells that the segment crosses, 
times the cost of accessing the data structure for the wide tetrahedra 
at each visited cell. Since we have allocated to each of the 
$O^*(D^{4j})$ cells at level $j$ the storage parameter $O(s/D^{4j})$,
the cost of accessing the structure for wide tetrahedra in a $j$-level
cell is, according to (\ref{eq:qns1}), at most
\[
{\small
Q_{\rm inner}(n,s) = O^*\left( \frac{n/D^{2j}}{(s/D^{4j})^{1/4}} + 
\left(\frac{(n/D^{2j})^{3/2}}{(s/D^{4j})^{1/2}}\right)^{5/6} \right) = 
O\left( \frac{n}{ D^{j} s^{1/4} } +
\frac{n^{5/4}}{D^{5k/6}s^{5/12}} \right) . }
\]
Summing over all $j$-level cells, for all $j$, and then 
adding the bottom-level cost, and the cost of traversing 
the structure with the query segment, the overall cost of a query is
(we remind the reader that so far we only consider the case where $s\le n^2$)
\begin{equation}
  \label{qtrade}
  O^*\left( D^{k} + \frac{n^{5/4} D^{k/6} }{s^{5/12}} + \frac{n}{s^{1/4}} + \frac{n}{s^{1/6}D^{k/3}} \right) .
\end{equation}
We choose $k$ to (roughly) balance the second and the last terms; specifically, we choose
\[
D^k = \sqrt{\frac{s}{n}} .
\]
Since $D^{k}$ should not exceed $O^*(n^{1/2})$, we require for this choice of $k$
that $s = O^*(n^{2})$. In this case it is easily verified that the second and last terms, 
which are $O^*(n^{7/6}/s^{1/3})$, dominate both the first and third terms (recall that we 
assume $s \ge n$), and the query time is therefore
\[
O^*(n^{7/6}/s^{1/3}) .
\]
For larger values of $s$, that is, when $s = \Omega^*(n^{2})$ (but we still
assume $s\le n^3$), we balance the first term with the last term, so we choose
\[
D^k = O^*\left(\frac{n^{3/4}}{s^{1/8}}\right) .
\]
Note that in this range we indeed have that $D^{k} = O^*(n^{1/2})$.
Moreover, in this case the first and last terms dominate the second and 
third terms, as is easily verified. Therefore the query time is
\[
O^*(n^{3/4}/s^{1/8}) .
\]
As already promised, the case where the query segment lies on the zero set 
in the current subproblem will be presented later.

\paragraph{Handling the range $n^3 < s\le n^6$.}
It remains to handle the range $n^3 < s\le n^6$.
Informally, at each cell $\tau$ of the polynomial partition, at any level
$j$ of the $D$-recursion, we have $n_\tau\le n/D^{2j}$ wide tetrahedra
and storage parameter $s_\tau = s/D^{3j}$. Since $s\ge n^3$, we also have
$s_\tau \ge n_\tau^3$. With such `abundance' of storage, we run the 
$r_0$-recursion until we reach subproblems of constant size, in which 
case we simply store the list of wide tetrahedra at each bottom-level node,
and the query simply inspects all of them, at a constant cost per subproblem. Hence
the cost of a query at $\tau$ is $O^*(n_\tau/s_\tau^{1/4})$.
To be precise, this is the case as long as $s_\tau \le n_\tau^4$.
If $n^3\le s \le n^4$ there will be some level $j$ of the $D$-recursion at whose cells $\tau$
$s_\tau = s/D^{4j}$ becomes larger than $(n/D^{2j})^4 \ge n_\tau^4$, and then 
the cost becomes $O^*(1)$. When $n^4 < s \le n^6$ the cost becomes $O^*(1)$ 
right away (and stays so). That is, the cost 
of a query in the structure for wide tetrahedra at a cell $\tau$ at level $j$ is
\begin{align*}
  O^*\left( \frac{(n/D^{2j})} {(s/D^{4j})^{1/4}} \right) = 
  O^*\left( \frac{n} {s^{1/4} D^{j}} \right) , & \qquad\text{for $s \le \frac{n^4}{D^{4j}}$} \\ 
  O^*\left(1\right) , & \qquad\text{for $s > \frac{n^4}{D^{4j}}$} .
\end{align*}
Since a query visits $O^*(D^{j})$ cells $\tau$ at level $j$, 
the overall cost of searching amid the wide tetrahedra, over all levels, 
is easily seen to be
\begin{align*}
  O^*\left( \frac{n} {s^{1/4}} \right) , & \qquad\text{for $n^3\le s \le n^4$} \\ 
  O^*\left( D^k \right) , & \qquad\text{for $n^4 < s \le n^6$} ,
\end{align*}
where $k$ is the depth of the $D$-recursion.

Querying amid the narrow tetrahedra is again done as in Section~\ref{sec:shoot} 
(once again, recall that we now consider the case where $s > n^3$, whereas earlier 
in this section we assumed $s \le n^3$). At each node $\tau$ at the bottom level 
$k$ of the $D$-recursion we use the data structure described in Section~\ref{app:trad},
which, with at most $n/D^{2k}$ narrow tetrahedra and storage parameter $s/D^{4k}$, 
answers a query in time 
\[
O^*\left( \frac{(n/D^{2k})} {(s/D^{4k})^{1/6}} \right) = 
O^*\left( \frac{n}{D^{4k/3}s^{1/6}} \right) .
\] 
We multiply by the number of cells that the query visits, namely $O^*(D^{k})$, 
and add the cost $O^*(D^{k})$ of traversing these cells, for a total of
\[
O^*\left( D^{k} + \frac{n}{s^{1/4}} + \frac{n}{D^{k/3}s^{1/6}} \right) .
\]
In other words, we get the same asymptotic bound as in (\ref{qtrade}), except for the
second term which is missing now (this term corresponds to querying at the
bottom-level nodes of the $r_0$-recursion on the wide tetrahedra, which
is not needed when $s > n^3$, since these bottom-level subproblems now have constant 
size). Repeating the same analysis as above, we get the same bound for the query cost.

\paragraph{Handling the zero set.}
% \esther{Here I only give a sketch, it is a repetition/easy variant of the calculations we have done above.}

The analysis for the zero set is done similarly to the analysis presented earlier 
in this paper, and to the one in~\cite{trishoot}, and is quite straightforward.
We do not provide a full description of these details, but only highlight the 
differences, from which we conclude that the query time bound is subsumed by 
that obtained when $\rho$ does not lie on the zero set.

Specifically, let us consider the query time bound obtained for the wide tetrahedra 
in~(\ref{eq:qns_0}). This bound also subsumes the bounds obtained for the case where 
$\rho^*$ is contained in the zero set $Z(G)$ of the second partitioning polynomial.
The bound holds for $n \le s \le n^2$, and for larger values of $s$ it becomes 
$O^*(n/s^{1/4})$, as long as $n^2 \le s \le n^4$, and $O^*(1)$ for $n^4 < s \le n^6$. 
We note that at every level $j$ of the recursion on the narrow tetrahedra we allocate 
to each subproblem the storage parameter $s/D^{3j}$, and the bound on the number of 
(wide and narrow) tetrahedra is still $O(n/D^{2j})$. Therefore at the bottom level 
$k$ we obtain an overall query time of
\[
O^*\left(D^{k} + \frac{n^{5/3} D^{k/6}}{s^{5/6}} + \frac{n}{s^{1/4} D^{k/4}} + \frac{n}{s^{1/6} D^{3k/2}}\right) .
\]
This bound is subsumed by the bound in~(\ref{qtrade}), for $s \ge n$, as is easily verified.
Therefore adding the query time for segment intersection within
$Z(f)$ does not increase the asymptotic bound in~(\ref{qtrade}).

     We next analyze the case where the query segment lies on the zero set.
     In order to obtain the trade-off bounds for segment intersection within $Z(f)$,
     we recall the multi-level data structure presented in Section~\ref{sec:onzf}. 
     Each level in this data structure is either a one- or a two-dimensional search tree, 
     where the dominating levels are those where we need to apply a planar decomposition 
     over a set of planar regions (or in an arrangement of algebraic arcs) and preprocess it
     into a structure that supports point-location queries. A standard property of multi-level 
     range searching data structures is that the overall complexity of their storage (resp.,
     query time) is governed by the level with dominating storage (resp., query time) bound,
     up to a polylogarithmic factor~\cite{AE99}. Recall that in each level of our data structure 
     we form a collection of canonical sets of the arcs in $\Gamma$, which are passed on to the 
     next level for further processing. Our approach is to keep forming these canonical sets, 
     where at the very last level we apply the segment-intersection data structure of Pellegrini~\cite{Pel}, 
     as described above. Therefore the overall query cost (resp., storage and preprocessing complexity) 
     is the sum of the query (resp., storage and preprocessing time) bounds over all canonical sets 
     of arcs that the query reaches (resp., all the sets) at the last level.
     
     We now sketch the analysis in more detail. In order to simplify the presentation, we consider 
     one of the dominating levels, and describe the segment-intersection data structure at that level. 
     As stated above, we build this data structure only at the very last level, but the analysis for
     the dominating level subsumes the bounds for the last level, and thus for the entire multi-level 
     data structure, up to a polylogarithmic factor. In such a scenario we have a set of algebraic 
     arcs (or graphs of functions, or semi-algebraic regions represented by their bounding arcs),
     which we need to preprocess for planar point location. This is done using the technique of 
     $(1/r)$-cuttings (see~\cite{CF-90}), which forms a decomposition of the plane into $O(r^2)$ 
     pseudo-trapezoidal cells, each meeting at most $n/r$ arcs (the ``conflict list'' of the cell). 
     The overall storage complexity is thus $O(nr)$. More precisely, to achieve preprocessing time 
     close to $O(nr)$, one needs to use so-called \emph{hierarchical-cuttings} (see~\cite{Mat} 
     and also \cite{AES}), in which we construct
     a hierarchy of cuttings using a constant value $r_0$ as the cutting parameter, instead of the
     nonconstant $r$. Using this approach, both storage and preprocessing cost are $O^*(nr)$. 
     Let $s$ be our storage parameter as above, so we want to choose $r$ such that $s = r n$.
     Thus we obtain that each cell of the cutting meets at most $n^2/s$ arcs.
     Following our approach above, for each cell of the cutting, the amount of allocated storage is $s/r^2 = n^2/s$.
     We are now ready to apply Pellegrini's data structure, leading to a query time of $O^*\left(\frac{n^{3/2}}{s^{3/4}}\right)$.
     Integrating this bound into the query time in~(\ref{qtrade}), we recall that at each level $0 \le j \le k$ the actual
     storage parameter is $O(s/D^{3j})$, and the number of tetrahedra at hand is $O(n/D^{2j})$. 
     We now need to sum the query bound over all $O(D^j)$ cells reached by the query at the $j$th level, 
     and over all $j$. We thus obtain an overall bound of
     $$
     O^*\left(D^k \frac{(n/D^{2k})^{3/2}}{(s/D^{3k})^{3/4}} \right) =
     O^*\left(\frac{n^{3/2} D^{k/4} }{s^{3/4}} \right) .
     $$
     This is exactly the second term in~(\ref{qtrade}). Therefore adding the query time 
     for segment intersection within $Z(f)$ does not increase the asymptotic bound in~(\ref{qtrade}).
     
     We comment that the overall storage and preprocessing time is $O^*(s)$ (see our discussion below).
     We also comment that the query bound we obtained applies when $n \le s \le n^2$. When $s$ exceeds $n^2$, 
     every cell of the cutting has a conflict list of $O(1)$ elements, which the query can handle in 
     brute-force. This immediately brings the query time, for queries on the zero set, to $O^*(1)$.

\paragraph{Wrapping up.}
In summary, our analysis implies that the query bound $Q(n,s)$ satisfies:
\begin{equation}
  \label{eq:trade-q}
  Q(n,s) = \begin{cases}
    O^*\left( \frac{n^{7/6}}{s^{1/3}} \right) , & s = O^*(n^{2}) , \\
    O^*\left( \frac{n^{3/4}}{s^{1/8}} \right) , & s = \Omega^*(n^{2}) .
  \end{cases}
\end{equation}
The overall storage (and expected preprocessing) is $O^*(s)$. Indeed,
we allocate to each subproblem, at any level $j$, the storage parameter $s/D^{4j}$,
so at each fixed level the total storage (and expected preprocessing) complexity is $O^*(s)$.
Since there are only logarithmically many levels, the overall storage (and expected preprocessing)
is $O^*(s)$ as well. This completes the proof of Theorem~\ref{thm:tradeoff}.

Note that for the threshold $s = n^{2}$, both bounds yield a query cost of $O^*(n^{1/2})$.
Note also that in the extreme cases $s = n^6$, $s = n$
(extreme for the `six-dimensional' tradeoff mentioned in Section~\ref{app:trad}), we get the 
respective bounds $O^*(1)$ and $O^*(n^{5/6})$ for the query time.
In this case, when either $s = n$ or $s = n^6$ 
we have $D^{k} = O(1)$, implying that we handle all the narrow tetrahedra at the root of the recursion tree.
That is, we use the technique of Section~\ref{app:trad} 
only once. Informally, the bound in~(\ref{eq:trade-q})
`pinches' the tradeoff curve and pushes it down. The closer $s$ is to 
$\Theta(n^{2})$, the more significant is the improvement.
See Figure~\ref{fig:trade}.

%-------------------------------------
\paragraph{Processing $m$ queries.}
The improved tradeoff in (\ref{eq:trade-q})
implies that the overall expected cost of processing $m$ queries with $n$ input 
tetrahedra, including (expected) preprocessing cost, is 
\[
O^*(s + m Q(n,s)) =
\begin{cases}
O^*\left( s + \frac{m n^{7/6}}{s^{1/3}} \right) , & s = O^*(n^{2}) , \\
O^*\left( s + \frac{m n^{3/4}}{s^{1/8}} \right) , & s = \Omega^*(n^{2}) .
\end{cases}
\]
To balance the terms in the first case we choose $s = m^{3/4}n^{7/8}$.
This choice satisfies $s = O^*(n^{2})$ when $m\le n^{3/2}$.
To balance the terms in the second case we choose $s = m^{8/9}n^{2/3}$.
This choice satisfies $s = \Omega^*(n^{2})$ when $m\ge n^{3/2}$.
Recall also that $s$ has to be in the range between $n$ and $n^6$. 
So in the first case we must have $m^{3/4}n^{7/8} \ge n$, or $m\ge n^{1/6}$.
Similarly, in the second case we must have $m^{8/9}n^{2/3} \le n^6$, or $m\le n^{6}$.
We adjust the bounds, allowing also values of $m$ outside this range, by adding the
near-linear terms $O^*(n)$ and $O^*(m)$, respectively, which dominate the bound 
for such off-range values of $m$. This establishes Corollary~\ref{cor:queries}.
% %-------------------------------
% \begin{corollary}
  % \label{cor:queries}
  % We can process $m$ segment-intersection or ray-shooting queries on $n$ 
  % tetrahedra so that the total expected cost is
  % \begin{equation}
    % \label{eq:trade-queries}
    % \max\Bigl\{ O^*(m^{3/4}n^{7/8} + n),\; O^*(m^{8/9}n^{2/3} + m) \Bigr\} .
  % \end{equation}
% \end{corollary}
% %-------------------------------

% \micha{We also need to state, earlier in the section, the tradeoff itself in a theorem.}

%-------------------------------------------
\section{Output-Sensitive Construction of Arrangements of Tetrahedra and of Intersections of Polyhedra in $\reals^4$}
\label{sec:arr}

The results of Section~\ref{sec:bip} can be applied to construct the arrangement $\A(\T)$ 
of a set $\T$ of $n$ tetrahedra in $\reals^4$ in an output-sensitive manner. 
A complete discrete representation of $\A(\T)$ requires, at the least, the collection of
all faces, of all dimensions, of the arrangement, and their adjacency structure.
Concretely, for each $j$-dimensional face $\varphi$, for $j=0,1,2,3$, we want the 
set of all $(j+1)$-dimensional faces that have $\varphi$ on their boundary. 
Conversely, for each $j$-dimensional face $\varphi$, for $j=1,2,3,4$, we want the 
set of all $(j-1)$-dimensional faces that appear on $\bd\varphi$.

We begin by considering the task of computing all the nonempty intersections of 
pairs, triples, and quadruples of tetrahedra of $\T$. This will yield the set of
vertices, and provide an infrastructure for computing the $j$-faces, for $j=1,2,3$.
Denote the number of these intersections as $k_2$, $k_3$, and $k_4$, respectively.
Note that we always have $k_4 \ge k_3\ge k_2$.

To simplify the description we assume that the tetrahedra are in general position,
although a suitable adaptation of the following machinery can handle degenerate cases too.

\paragraph{Reporting pairwise intersections.}
Two tetrahedra in general position in $\reals^4$ intersect in a two-dimensional 
convex polygon of constant complexity, and it suffices to report one vertex of 
each nonempty polygon, in order to detect all intersecting pairs of tetrahedra. 
As is easily checked, such a vertex is either an
intersection of an edge of one tetrahedron with the other tetrahedron, or an
intersection of two 2-faces (triangles), one from each tetrahedron.

Reporting vertices of the first kind (edge-tetrahedron intersections) can be 
done using the machinery in Theorem~\ref{thm:standard}, whose details are provided
in Section~\ref{app:trad}, which takes $O^*(n^{12/7} + k_2)$ time.\footnote{%
  Although this part can be performed faster, as described in Section~\ref{sec:shoot},
  %(and in Section~\ref{app:trade}),
  we use the standard solution, since we do not
  have a similar improvement for the construction of vertices of the second kind.}
Reporting vertices of the second kind 
(triangle-triangle intersections) is done using the machinery in 
Section~\ref{sec:bip}, which also takes $O^*(n^{12/7} + k_2)$ time.
% \micha{For ourselves: if an edge $e$ of one tetrahedron crosses a second tetrahedron 
  % $T'$, this does not mean that $e$ crosses a 2-face of $T'$!}
% \esther{Right, $e$ can intersect the hyperplane containing $T$.}

\paragraph{Reporting triple and quadruple intersections.}
We iterate over the input tetrahedra. For each fixed tetrahedron $T_0$,
the previous step provides us with all the other tetrahedra that intersect
$T_0$. Denote their number as $k_{T_0}$, and observe that
$\sum_{T_0} k_{T_0} = 2k_2$. We form the nonempty intersections
$T_0\cap T$, and triangulate each of them. We obtain a collection
of $O(k_{T_0})$ triangles, all contained in ($T_0$ and therefore 
also in) the hyperplane $h_{T_0}$ supporting $T_0$. 

We have thus reduced our problem to that of reporting all pairwise and 
triple intersections in a set of $m = O(k_{T_0})$ triangles in $\reals^3$.
This can be solved using the algorithm in~\cite{trishoot}, by a procedure
that runs in $O^*(m^{3/2} + \ell_{T_0})$ time, where $\ell_{T_0}$ is
the number of triple intersections of the triangles. Note that 
$\sum_{T_0} \ell_{T_0} = O(k_4)$. 

Adding up this cost over all tetrahedra $T_0$, the overall running time is
\[
O^*\left( \sum_T k_T^{3/2} + k_4 \right) =
O^*\left( n^{1/2} \sum_T k_T + k_4 \right) = O^*(n^{1/2}k_2 + k_4) .
\]

\paragraph{Constructing the arrangement.}
For each tetrahedron $T_0$, it is fairly routine to obtain, from the
information collected so far, the full three-dimensional arrangement 
within $T_0$, using standard techniques in three dimensions; we omit
here these standard details. This gives us all the $j$-faces of the
four-dimensional arrangement $\A$, for $j=0,1,2,3$, and their adjacency 
information. The local adjacency information in $\reals^4$ is also 
available from this data. By local adjacency we mean the adjacency 
between a $j$-face and the $j'$-faces on its boundary, for $j' < j$, 
over all such pairs of faces. For completion we need to identify 
disconnected pieces of the boundary of each four-dimensional cell, 
and record their adjacency to that cell. This can be done by $x_4$-vertical 
ray shooting from the $x_4$-highest point of each connected three-dimensional 
complex of faces. This calls for performing $O(n)$ $x_4$-vertical ray 
shooting queries in a set of $n$ tetrahedra in $\reals^4$, which can 
be done using the machinery presented in Theorem~\ref{thm:standard},
or by an even simpler mechanism (since all the rays are vertical).

We have thus established the bound stated in Theorem~\ref{thm:arr}.

%------------------------------------
\paragraph{Output-sensitive construction of the intersection or union of polyhedra in $\reals^4$.}
As another application, consider the problem where we have two not necessarily convex
polyhedra $R$ and $B$ in $\reals^4$, whose boundaries consist of, or can be triangulated into
$O(n)$ faces of all dimensions, which are segments, triangles, and tetrahedra. 
The goal is to construct their intersection $R\cap B$ in an output-sensitive manner;
a similar application has been shown in~\cite{trishoot} for the three-dimensional problem.
We note that computing the union $B\cup R$ can be done using a very similar approach, within 
the same asymptotic time bound.

In order to compute $R\cap B$, we first apply the above algorithm to construct, in 
an output-sensitive manner, the arrangement $\A(R \cup B)$ of the two polyhedra $R$ and $B$
(specifically, we build the arrangement of the tetrahedra comprising the boundaries of $B$ and $R$).
We then label each cell (of any dimension) of $\A(R \cup B)$ with the appropriate
Boolean operation, that is, whether it either lies in $R \setminus B$, $B \setminus R$,
$B \cap R$, or in the complement of $B \cup R$.
Collecting all the cells of the desired kind (e.g., those in $B\cap R$), and computing the adjacency relation between them, we obtain a suitable representation of the intersection. This establishes the bound stated in Theorem~\ref{thm:arr}(ii).

%Consider the construction of the vertices of $R\cap B$.
%Any new vertex of $\bd(R\cap B)$ is either an intersection of either
%(i) an edge of $R$ with a tetrahedral facet of $B$, or
%(ii) a triangular 2-face of $R$ with a triangular 2-face of $B$, or
%(iii) a tetrahedral facet of $R$ with an edge of $B$.
%All these intersections can be computed using the machinery developed in this paper,
%and take a total of $O^*(n^{12/7}+k)$ time, where $k$ is the output size. 
%Finding the other new features of the intersection boundary is also easy to perform. 
We comment that extending the analysis to the intersection of more than two (albeit, still a constant number of)
input polyhedra can also be done, following the same machinery as in the construction of an arrangement
of tetrahedra, as presented above. It is easy to verify that in this case we obtain the same asymptotic
bound stated in Theorem~\ref{thm:arr}(ii).

%---------------------------------------------------------------------
\section{Detecting or Reporting Intersections between 2-Flats and Lines in $\reals^4$}
\label{app:int}

As a final application of our machinery, we consider the problem where we are given a 
set $R$ of $n$ red 2-flats and a set $B$ of $n$ blue lines in $\reals^4$, and 
the detection problem asks whether there exists a pair of intersecting 
objects in $R\times B$. In the reporting version we want to report all such pairs.
We only consider the batched version of the problem, but a similar approach can 
also handle the preprocessing-and-query variant.

We solve the detection problem by regarding the problem as a special degenerate 
(and much simpler) instance of the segment intersection setup (and also of the
triangle-triangle intersection setup), in which we regard the, 
say, red 2-flats as degenerate tetrahedra (unbounded and 
of zero volume), construct the data structure of Section~\ref{sec:shoot},
and query it with each of the blue lines. There exists a red-blue
pair of intersecting objects if and only if at least one query has
a positive outcome---the corresponding blue query line hits a red 2-flat.
Using the bounds and notation given in Corollary~\ref{cor:queries}, this can be performed in expected time
$\max \{ O^*(m^{3/4}n^{7/8} + n),\; O^*(m^{8/9}n^{2/3} + m)\}$, and since $m=n$ in our case this bound is
$O^*(m^{13/8})$. This is a clear improvement over the bound $O^*(n^{12/7})$ obtained using the initial 
approach presented in Section~\ref{app:trad}. Indeed, in this latter approach, 
with storage parameter $s$, a query takes $O^*(n/s^{1/6})$ time, and thus $n$ queries cost
$O^*(n^2/s^{1/6})$ time. Balancing these costs results in overall expected running time of $O^*(n^{12/7})$.
Similar improvements are obtained for other values of $m$, as long as $n^{1/6}\ll m\ll n^6$
(see the analysis in the preceding section).

Since there are no wide tetrahedra in this special variant, there is
no need to construct the auxiliary data structure for wide tetrahedra,
as in Section~\ref{sec:shoot}, and we simply construct the recursive
hierarchy of polynomial partitions, where each cell in each subproblem 
is associated with the set of red 2-flats that cross it. A blue query line
$\ell$ is propagated through the cells that it crosses until it either comes
to lie on the zero set of the current partitioning polynomial,
%(a case considered next),
or reaches bottom-level cells, and we check, in each such cell, whether $\ell$
intersects any of the $O(1)$ red 2-flats associated with the cell.

An easy adaptation of our machinery allows us to report all $k$ red-blue
intersecting pairs in expected time $O^*(n^{13/8} + k)$.

%Handling lines that lie fully in the zero set $Z(f)$ is also an easy  task, which can be performed using the planar segment-intersection range  searching presented in Section~\ref{sec:onzf}, which also supports counting queries, as is easily checked; the further easy details are omitted.

%Both correctness and runtime analysis follow easily, as special
%and simpler instances of the analysis in Section~\ref{sec:shoot}.
%Note that here we do not face the issue of non-disjointness of canonical sets
%of wide triangles (because there are no wide triangles), 
%which has prevented us from extending the technique
%to segment-triangle intersection counting problems; see Section~\ref{sec:seg}.

In summary, we have:
%-------------------------------
\begin{theorem} 
Given $n$ blue lines and $n$ red 2-flats in $\reals^4$, one can detect whether
some blue line intersects some red 2-flat in $O^*(n^{13/8})$ expected time. One can also
report all $k$ red-blue intersections in $O^*(n^{13/8} + k)$ expected time. 
\end{theorem}
%-------------------------------

We remark that this case can also be considered as a special case of the triangle-triangle intersection setup.

\bibliography{ray4z}

\end{document}